\documentclass[%
 reprint, onecolumn,
 amsmath, amssymb,
 aps,
]{revtex4-2}

\usepackage{graphicx}
\usepackage{dcolumn}
\usepackage{bm}
\usepackage{xcolor}
\usepackage{amsmath}
\usepackage{mathrsfs}
\usepackage{multirow}

\usepackage{hyperref}
\usepackage{subeqnarray}

\begin{document}
\preprint{AIP/123-QED}
\title{Electrostatic effects on settling cloud droplets: from pair collisions to size distribution evolution}

\author{Pijush Patra}
\affiliation{Nordita, KTH Royal Institute of Technology, and Stockholm University, Stockholm 10691, Sweden}
\author{J. S. Wettlaufer}
\homepage{Author to whom correspondence should be addressed: jw@fysik.su.se, john.wettlaufer@yale.edu}
\affiliation{Departments of Earth \& Planetary Sciences,  Mathematics, and Physics, Yale University, New Haven, Connecticut 06520-8109, USA}
\affiliation{Nordita, KTH Royal Institute of Technology, and Stockholm University, Stockholm 10691, Sweden}

\begin{abstract}
We investigate how droplet charge and a uniform vertical electric field influence collisions between cloud droplets settling in still air and how these collisions shape the droplet size distribution. We compute collision efficiencies by analyzing the relative trajectories of pairs of spherical, conducting droplets, accounting for droplet inertia, exact hydrodynamic, electrostatic, and van der Waals forces. The hydrodynamic interactions are based on exact two-body solutions for Stokes flow with interfacial mobility and internal circulation. In the absence of an applied electric field, collisions between like-charged droplets are suppressed, whereas collisions between oppositely-charged droplets are enhanced. A vertical electric field promotes collisions between uncharged droplets through induced polarization and, when sufficiently strong, can overcome the repulsion between like-charged droplets. When the larger droplet has a radius of $10$ or $20\,\text{\textmu}\mathrm{m}$, raising the electric field strength from $5\times10^{4}$ to $3\times10^{5}\,\mathrm{V\,m^{-1}}$ can enhance the collision efficiency of uncharged droplet pairs by an order of magnitude. Depending on droplet size and field strength, for oppositely charged pairs, an applied electric field can either enhance or suppress collisions relative to the zero-field case. Using collision kernels computed from droplet pair trajectories, we solve the stochastic collection equation with a flux method on a two-dimensional grid that resolves droplet mass and charge. For an initial mean droplet radius of $10\,\text{\textmu}\mathrm{m}$, by $40\,\mathrm{min}$ in every electrified case the large-droplet mode of the mass spectrum reaches a larger radius than in the hydrodynamic reference case. These findings suggest that electrostatic interactions can hasten warm rain initiation by accelerating droplet growth through collision and coalescence.
\end{abstract}

\maketitle

\section{Introduction} \label{sec:Introduction}

The increasing frequency of extreme weather events, coupled with anthropogenic climate change, poses a significant global challenge that has captured a great deal of recent attention \citep{Myhre2019ExtremePrecipitation}. 
Clouds play a vital role in the Earth's climate system by regulating precipitation, governing hydrological cycles, and influencing the Earth's radiative balance \citep{bony2015clouds}. The effects of clouds on atmospheric processes operate across a wide range of temporal and spatial scales. While visible clouds can span distances of order a few kilometers, individual water droplets are typically only a few microns in size. Consequently, large-scale models, such as numerical weather prediction (NWP) simulations, cannot capture the processes occurring at smaller scales. Therefore, clouds are a major source of uncertainty in climate models, primarily due to the complexities involved in accurately parameterizing how small-scale processes impact macroscale phenomena \citep{schneider2017climate}. Microphysical processes are fundamental in determining the lifetime, radiative properties, and precipitation efficiencies of clouds. Thus, to improve our understanding and prediction of precipitation patterns, the Earth's energy budget, climate sensitivity, and the influence of clouds on human activities, 
understanding these microphysical processes constitutes a major research thrust. These processes involve the activation of cloud condensation nuclei (CCN), condensational growth of droplets, and collision-coalescence between droplets.

Understanding the underlying mechanisms that lead to warm rain initiation within approximately thirty minutes remains a critical challenge in cloud microphysics \citep{grabowski2013growth,poydenot2024pathways}. Various mechanisms have been proposed to explain this phenomenon, including the influence of giant CCN \citep{yin2000effects,barahona2010comprehensively}, the entrainment and mixing of unsaturated air with the cloud \citep{kumar2013cloud,yeom2025cloud}, turbulent fluctuations in water vapor supersaturation \citep{sardina2015continuous}, and turbulent collision-coalescence \citep{sundaram1997collision,chen2018turbulence}. Despite these efforts, the rapid broadening of the droplet size distribution (DSD) remains largely unresolved. In situ measurements (see \citet{prabha2011microphysics}) have shown broad DSDs in warm clouds, which are unlikely to result from condensational droplet growth alone. Therefore, the primary drivers of rain formation in warm clouds are collisions and subsequent coalescence between droplets, with the evolution of the DSD strongly influenced by the collision rate. While much of the research has predominantly concentrated on the roles of gravity and turbulence on droplet-droplet collisions (see \citet{shaw2003particle,devenish2012droplet,grabowski2013growth,ababaei2023collision} and references therein), relatively few studies have explored the impact of electrostatic forces in collision dynamics \citep{schlamp1976numerical,tinsley2000effects,khain2004rain,lu2015charged,guo2021enhancement,patra2023collision,dubey2024critical,patra2025gravity,patra2025electric}. An accurate description of the effects of electrostatic interactions in droplet collisions is crucial for developing a more realistic parameterization for NWP models. Furthermore, electrostatic forces may be crucial in triggering collisions among small droplets, thereby introducing the necessary size disparities for continued droplet growth. Here we focus on the influence of droplet electric charges and an external electric field on the collision dynamics of cloud droplets settling in still air.

In weakly electrified clouds, droplets acquire electrical charges primarily through the diffusion of atmospheric ions produced by cosmic rays \citep{pruppacher1997microphysics,wang2013physics}. A ubiquitous downward fair-weather electric field exists due to the potential difference between the Earth's surface and the ionosphere. During fair-weather conditions, the strength of these electric fields in clouds typically ranges from $10^2$ to $10^3$ V/m \citep{gunn1948electric}. The mobilities of atmospheric ions in this fair-weather electric field generate a downward ionosphere–Earth current density on the order of $10^{-12}$ Am$^{-2}$ \citep{rycroft2008overview}. When the fair-weather current passes through a horizontally extensive stratiform cloud, the conductivity transition between the clear air and cloudy regions results in a net accumulation of positive space charges at the upper edge and negative space charges at the lower edge of the cloud \citep{nicoll2016stratiform}. Consequently, droplets at the top of the cloud gain positive charges, while those at the bottom acquire negative charges. \citet{zhou2007production} gave a theoretical estimate of the magnitude of charges on droplets with a mean radius of $10$ \textmu m ranges from $50e - 100e$ ($e=1.602 \times 10^{-19}$ C represents the elementary charge), which is consistent with observations of \citet{beard2004aircraft}. Based on observations, various empirical relationships between droplet radii and mean absolute charges on droplets have been documented in the literature (see Chap. 18.4 in \citet{pruppacher1997microphysics}). For this work, we will use the quadratic relation $q = c \times 3.33 \times 10^{-6} a^2$, where $q$ is the mean absolute droplet charge in coulombs (C), $a$ is the droplet radius in meters, and $c$ is an empirical parameter. The values of $c$ ranging from $0.1$ to $0.6$ correspond to weakly electrified clouds \citep{zhang2024preliminary}.

Unlike in weakly electrified clouds, in thunderclouds, strong electric fields develop due to cloud charging, which facilitates charge separation. Charge separation involves several mechanisms, which can be broadly classified into convective, inductive, and non-inductive processes. These mechanisms are comprehensively discussed in chapter 18.5 of \citet{pruppacher1997microphysics} and chapter 14.4 of \citet{wang2013physics}. Among the various theories proposed to explain cloud charging, the most extensively studied is the charging that occurs when ice particles collide and rebound. Laboratory experiments have demonstrated the importance of collisional charging during ice-ice collisions in thundercloud electrification \citep{mason2000charge,saunders2006laboratory}, which has inspired the development of theoretical models to explain the phenomenon \citep{dash2001theory,dash2003surface,jungwirth2005possible}. Electric field strength in moderately to highly electrified clouds is in the range of $10^4-10^5 \text{V/m}$ \citep{gunn1948electric,pruppacher1997microphysics,rakov2003lightning}. Under average conditions for strongly electrified clouds, the parameter $c$ is approximately $2$, with an upper limit of around $7$ attributable to conduction charging (see chapter 18.4 of \citet{pruppacher1997microphysics}). Notably, the charge on a droplet calculated from this relationship remains below the maximum charge the droplet can sustain, as determined by phenomena such as electromechanical instability \citep{rayleigh1882xx} or corona discharge \citep{meek_electrical_breakdown_1954}.

To quantify the impacts of surface charges and electric fields on collision dynamics, it is crucial to accurately describe the electrostatic forces acting on droplets. A pioneering study by \citet{davis1964two} established a comprehensive theoretical framework for calculating the electrostatic forces between two charged conducting spheres subjected to an external electric field. This foundational work has served as the basis for subsequent investigations into electrically mediated droplet collisions that are relevant in cloud microphysics. Utilizing the work of \citet{davis1964two}, \citet{schlamp1976numerical}, \citet{klimin1994collision}, and more recently \citet{guo2021enhancement} and \citet{zhang2025collision}, computed collision efficiencies of two sedimenting charged droplets in the presence of external vertical electric fields. These studies have shown that electrostatic forces can significantly enhance the collision efficiencies between small droplets in strongly electrified clouds. In their cloud chamber experiment, \citet{wang2020electrostatic} reported that a strong electric field can cause significant collisional growth of droplets larger than $10$ \textmu m in diameter. In scenarios devoid of an external electric field, \citet{tinsley2000effects} and \citet{tinsley2006changes} accounted for the image charge forces to evaluate collision efficiencies of charged aerosol particles and droplets in weakly electrified clouds. Their findings revealed that electrical effects can lead to a considerable enhancement in collision efficiency, facilitating the scavenging of aerosol particles by droplets, even when the charges involved are of the same polarity. Similar calculations were performed by \citet{zhou2009scavenging}, using the electrostatic interaction force expression derived by \citet{davis1964two}. Laboratory experiments conducted by \citet{ardon2015laboratory} reported results that agreed with the theoretically calculated collision efficiency in the context of electro-scavenging. \citet{khain2004rain} generalized the formulation of image charge forces for calculating the collision efficiency of two charged conducting spheres. They demonstrated that such forces can significantly enhance droplet collision efficiency, and proposed that charged droplet seeding could be a potential technique for rain enhancement and fog elimination. Finally, A few experimental studies have also shown that electrostatic forces can significantly enhance the collision efficiency of charged droplets (see \citet{abbott1975charged,beard2002coalescence,magnusson2022collisions}).

The theoretical work cited above has not fully accounted for the hydrodynamic and electrostatic interactions that occur when the gap between the surfaces of two spherical droplets is much smaller than their radii. Ultimately, the fluid in this narrow gap must be squeezed out for two droplets to come into contact. Therefore, these close approach interactions are critical for accurately capturing the mechanisms underlying collision dynamics. Recent studies have highlighted the necessity of incorporating the exact lubrication forms of hydrodynamic and electrostatic interactions when analyzing the collision dynamics of pairs of rigid spheres settling under gravity \citep{patra2023collision,dubey2024critical,poydenot2024gap,patra2025gravity}. Here we investigate the influence of near-field behavior of hydrodynamic, electrostatic, and van der Waals interactions for cloud droplets sedimenting in still air, while considering the internal circulations and interfacial mobilities of droplets.

Consistent with the majority of previous studies of the collision dynamics of two sedimenting spheres (see, for example, \citet{rother2022gravitational,ababaei2023collision,dubey2024critical}), we assume that the inertia of the droplets is significant while the inertia of the surrounding fluid is negligible. This assumption enables us to use the Stokes equations to describe the hydrodynamic forces acting on the droplets. The sizes of the droplets under investigation are sufficiently large to neglect their Brownian motion. Additionally, we will treat cloud droplets as perfect conductors and disregard any possible deformation from their initial spherical shape during their motion. In \S \ref{sec:Methodology}, we will justify these assumptions in the context of cloud droplets settling in still air.

To understand the hydrodynamic interactions between spherical particles, researchers have developed several approximate methods, among which the superposition method is notable \citep{wang2005improved}. These methods are based on the solution to the Stokes flow generated by a single translating sphere. While they provide accurate estimates of the hydrodynamic forces between particles separated by large distances, they are inadequate for accurately predicting the hydrodynamic resistances--lubrication forces--encountered during the relative motion of two particles in near contact. Exact solutions to the Stokes equation for two hydrodynamically interacting rigid or fluid spheres are available in the literature (see \citet{ababaei2023collision} for a detailed review). Noteworthy contributions include those of \citet{haber1973low} and \citet{zinchenko1980slow}, who have derived exact solutions to the Stokes equation for two fluid droplets translating along and normal to their line of centers, respectively. We make use of these two studies to calculate the hydrodynamic forces acting on two fluid droplets settling under the influence of gravity.

In the lubrication regime, calculations of the hydrodynamic forces using the exact solutions described above can be computationally expensive. To address this issue, we use the lubrication forms of these forces. In addition to simplifying computations, these lubrication forms provide crucial mathematical insights into the mechanisms that govern collision dynamics. When two rigid spheres interact in a continuum fluid medium, the lubrication force scales as the inverse of the gap size between the spheres \citep[e.g.,][]{kim2013microhydrodynamics}. Therefore, to overcome the lubrication force and enable a collision, it is necessary to include a singular non-hydrodynamic attractive force, such as the van der Waals force (see \citet{davis1984rate}), or consider non-continuum effects \citep[e.g.,][]{dhanasekaran2021collision,patra2022brownian,patra2023collision}. In contrast, for spherical liquid droplets, due to the internal circulation and interfacial mobility, lubrication resistance is inversely proportional to the square root of the gap size, allowing collisions to occur within a finite time under an exerted constant force \citep{zinchenko1982calculation,davis1989lubrication}. Recently, \citet{rother2022gravitational} and \citet{ababaei2023collision} have utilized these solutions to compute the collision efficiency of a pair of unequal-sized fluid droplets settling under gravity and compared their findings with those obtained from the approximate model developed by \citet{wang2005improved}.

Accurate estimation of collision rates that incorporate detailed hydrodynamic and electrostatic interactions between droplets is crucial for realistic prediction of the DSD. Despite its significance, relatively few studies have examined the impact of electrostatic effects on the DSD evolution. A numerical study by \citet{khain2004rain} showed that strong droplet charging can substantially enhance collision rates, thereby accelerating raindrop formation in warm clouds. \citet{guo2021enhancement} demonstrated that in strongly electrified clouds, the electric charges on droplets and the atmospheric electric fields can enhance rain initiation. In the context of weakly electrified warm cumulus clouds, \citet{zhang2024preliminary} employed a Lagrangian particle-based cloud model to underscore that droplets with opposite charges can exert a significant influence on cloud evolution. Here we use more accurate droplet collision rates to investigate the evolution of the DSD under a wide range of electrical conditions.

We organize the remainder of the paper as follows. In \S \ref{sec:Methodology}, we outline the equations of motion and the forces acting on the droplets and describe the numerical procedure used to calculate efficiency through trajectory analysis. In \S \ref{sec:Relative_trajectory_and_collision_efficiency}, we present typical relative trajectories and results of the electrostatic effects on collision efficiency. Descriptions of the model setup for solving the stochastic collection equation (SCE) and results of the electrostatic effects on the evolution of the DSD are presented in \S \ref{sec:DSD_evolution}. Finally, we summarize our results and conclude in \S \ref{sec:Conclusions}.

\section{Methodology}\label{sec:Methodology}

\begin{figure}
\centering
\includegraphics[width=0.8\textwidth]{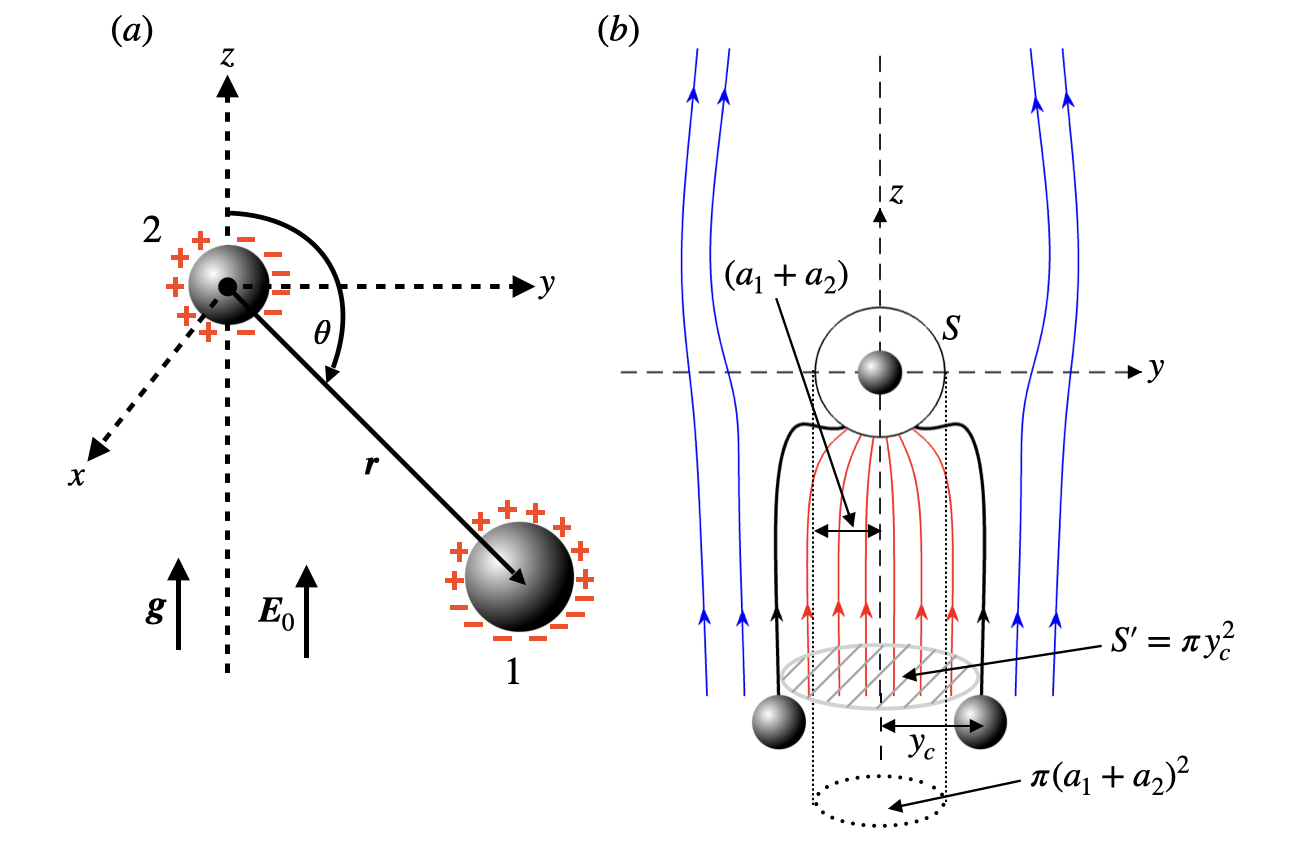}
\caption{(a) Schematic of two charged droplets subject to gravity ($\boldsymbol{g}$) and a vertical electric field ($\boldsymbol{E}_0$). Here, `1' represents the droplet with radius $a_1$ and charge $q_1$; `2' represents the droplet with radius $a_2$ and charge $q_2$. We use $\boldsymbol{\hat{e}}_r$ and $\boldsymbol{\hat{e}}_{\theta}$ as the unit vectors in $r$ and $\theta$ directions, respectively. (b) Schematic of typical relative trajectories and upstream collisional areas. The sphere marked `$\mathcal{S}$' denotes the projection of the collision sphere of radius $a_1+a_2$. The red, black, and blue lines represent colliding, limiting colliding, and open trajectories, respectively. The dotted circle, which has a radius of $a_1 + a_2$, denotes the upstream interception area for a pair of non-interacting droplets. The hatched circle, with a radius of $y_{c}$, represents the upstream interception area that defines the collision rate with interactions.}
\label{Schematic_binary_collisions_charged_droplets_with_gravity_electric_field}
\end{figure}

We consider two charged, conducting cloud droplets subjected to gravitational forces and a uniform vertical electric field (see figure \ref{Schematic_binary_collisions_charged_droplets_with_gravity_electric_field}$a$). In dilute systems such as clouds, where droplet volume fractions are on the order of $10^{-6}$ \citep{grabowski2013growth}, the probability that a third droplet substantially alters the relative motion of two interacting droplets is negligible. Therefore, we restrict our analysis to binary collisions. We define the characteristic length scale, denoted by $a^*$, and the velocity scale, denoted by $V^{(0)}_{12}$, as the mean radius of the two droplets and the relative velocity of two non-interacting droplets settling under gravity, respectively. The ratio of the characteristic length scale to the velocity scale defines the characteristic time scale, denoted by $t^*$. Mathematically, $a^*=(a_1+a_2)/2$, $V^{(0)}_{12}=|\boldsymbol{V}^{(0)}_{12}| = 2 C_v (\rho_d-\rho_f) \left(a^2_1-a^2_2\right)g/(9 \mu_f)$, and $t^* = a^*/V^{(0)}_{12}$, where $a_1$ is the radius of the larger droplet, $a_2$ is the radius of the smaller droplet, $\rho_d$ and $\rho_f$ are the densities of the droplet and the surrounding fluid, respectively, $g$ is the magnitude of the gravitational acceleration $\boldsymbol{g}$, and $\mu_f$ is the dynamic viscosity of the surrounding fluid. The factor $C_v$, which accounts for the internal circulation of viscous droplets, is given by $C_v=(\hat{\mu}+1)/(\hat{\mu}+(2/3))$, where $\hat{\mu}$ is the droplet-to-medium viscosity ratio (see example $4.3$ in \citet{kim2013microhydrodynamics}). The two essential parameters that characterize the relative geometry of the two spherical droplets are the size ratio, $\kappa = a_2/a_1$, and the center-to-center distance between the droplets $r$. The dimensionless surface-to-surface distance is defined as $\xi = [r-(a_1+a_2)]/a^* = (r/a^*)-2$, which we will use later in this section.

In this study, we ignore droplet deformations. The capillary number $Ca$, which measures the relative strength of hydrodynamic and capillary stresses, is defined as $Ca = \mu_f V_1^{(0)}/\sigma$, where $\sigma$ is the surface tension at the air-water interface. Here, $V_1^{(0)} = 2 C_v (\rho_d-\rho_f) g a_1^2/(3\mu_f)$ is the Hadamard–Rybczynski settling velocity of the larger droplet. To quantify the relative significance of electric stress in comparison to capillary stress, we define the electric capillary number $Ca_E = \epsilon_0 E_0^2 a_1/\sigma$, where $\epsilon_0 = 8.85 \times 10^{-12}$ Fm$^{-1}$ is the permittivity of air, and $E_0$ is the magnitude of the vertical electric field $\boldsymbol{E}_0$. These two dimensionless parameters effectively characterize the deformation behavior of the droplet. Let us consider a cloud droplet of radius $20$ \textmu m, with $\rho_d \approx 10^3$ kg m$^{-3}$, $\rho_f \approx 1.3$ kg m$^{-3}$, $\mu_f \approx 1.8 \times 10^{-5}$ Pa\,s, $\hat{\mu} \approx 10^2$, $E_0 = 10^5$ Vm$^{-1}$, and $\sigma \approx 72 \times 10^{-3}$ Nm$^{-1}$. These values give $Ca \approx 1.23 \times 10^{-5}$ and $Ca_E \approx 2.46 \times 10^{-5}$, both of which are sufficiently small to allow us to safely neglect droplet deformation. Furthermore, electrostatic and van der Waals forces may lead to impending collisions before the droplet surfaces come sufficiently close for pressure-induced and electric-field-induced deformations to become important. Nonetheless, these deformations near the contact region will alter the collision dynamics for larger droplets and strong electric fields, which we do not consider in this analysis.

The key dimensionless numbers that describe the interactions of two droplets settling under gravity are the droplet Reynolds number $Re_d$, the Stokes number $St$, and the P\'eclet number $Pe$. We define $Re_d$ based on the Hadamard–Rybczynski settling velocity and radius of the larger droplet as: $Re_d = \rho_f a_1 V_1^{(0)}/\mu_f = 2 C_v \rho_f(\rho_d-\rho_f) g a_1^3/(9\mu_f^2)$. The Stokes number serves as a measure of droplet inertia and is defined as: $St = \tau_d/t^* = 2C_v\rho_d(\rho_d-\rho_f) g a_1^3(1+\kappa)(1-\kappa^2)/(81\mu_f^2)$, where $\tau_d = 2\rho_d (a^*)^2/(9\mu_f)$ is the droplet relaxation time scale. The P\'eclet number, which captures the relative strengths of gravitational advection and Brownian diffusion, is given by: $Pe = a^* V^{(0)}_{12}/D^{(0)}_{12} = 2\pi (\rho_p-\rho_f)a_1^4 \kappa (1-\kappa^2)g/(3 k_B T)$, where $D_{12}^{(0)}=C_v k_B T (1+\kappa^{-1})/(6 \pi \mu_f a_1)$ is the relative Brownian diffusivity for two non-interacting droplets \citep{zinchenko1994gravity}. Here, $k_B=1.318 \times 10^{-23}$JK$^{-1}$ is Boltzmann's constant, and $T$ is the absolute temperature. Table~\ref{tab:sedimenting-droplet-parameters} gives the droplet Reynolds number $Re_d$, the Stokes number $St$, and the P\'eclet number $Pe$, for different size combinations of droplet pairs. Over the radius range $10\,\text{\textmu}\mathrm{m} \leq a_1 \leq 40\,\text{\textmu}\mathrm{m}$ of the larger droplet, $Re_d$ increases from $0.00876$ to $0.561$. We therefore use a creeping-flow approximation for the hydrodynamic interactions, while recognizing that corrections due to fluid inertia may become appreciable near the upper end of this range. For all size ratios tabulated in this radius range $Pe \gtrsim 10^3$, indicating that differential settling dominates Brownian diffusion on the scale of the mean droplet radius $a^*$. Accordingly, we neglect Brownian diffusion in the trajectory calculations. The Stokes number, however, depends strongly on both $a_1$ and $\kappa$. For example, when $a_1=20\,\mu\mathrm{m}$ and $\kappa$ increases from 0.40 to 0.99, $St$ decreases from 7.05 to 0.237,
as the isolated settling speeds of the two droplets approach one another. Since $St$ is of order unity or larger for many of the droplet pairs considered, we retain droplet inertia in the equations of motion.

\begin{table*}
\centering
\caption{Typical values of the droplet Reynolds number $Re_d$, the Stokes
number $St$, and the P\'eclet number $Pe$, for water droplets settling in air
at $T=275\,\mathrm{K}$. We use a water density of
$\rho_d=10^{3}\,\mathrm{kg\,m^{-3}}$, an air density of
$\rho_f=1.3\,\mathrm{kg\,m^{-3}}$, an air dynamic viscosity of
$\mu_f=1.8\times10^{-5}\,\mathrm{Pa\,s}$, a droplet-to-air dynamic
viscosity ratio of $\hat{\mu}=100$ (giving $C_v\simeq1.00331$), and a
gravitational acceleration of $g=9.81\,\mathrm{m\,s^{-2}}$.
Each two-row group corresponds to a fixed larger droplet radius $a_1$,
and the columns give the size ratio $\kappa=a_2/a_1$, where $a_2$ is the
radius of the smaller droplet. In rows labeled $Pe\,(\times 10^{-n})$,
multiply the tabulated entry by $10^n$ to obtain $Pe$.}
\label{tab:sedimenting-droplet-parameters}
\renewcommand{\arraystretch}{1.15}
\setlength{\tabcolsep}{3pt}
\begin{tabular}{@{}*{12}{c}@{}}
\hline
\hline
 & & & \multicolumn{9}{c}{$\kappa=a_2/a_1$} \\
\cline{4-12}
$a_1$ ($\text{\textmu}\mathrm{m}$) & $Re_d$ & Parameter
 & $0.05$ & $0.10$ & $0.20$ & $0.40$ & $0.60$
 & $0.80$ & $0.90$ & $0.95$ & $0.99$ \\
\hline

\multirow[c]{2}{*}{$5$} & \multirow[c]{2}{*}{$0.00110$} & $St$
 & $0.0981$ & $0.102$ & $0.108$ & $0.110$ & $0.0959$
 & $0.0607$ & $0.0338$ & $0.0178$ & $0.00371$ \\
 & & $Pe\,(\times 10^{-3})$
 & $0.168$ & $0.334$ & $0.649$ & $1.13$ & $1.30$
 & $0.973$ & $0.578$ & $0.313$ & $0.0665$ \\
\hline

\multirow[c]{2}{*}{$10$} & \multirow[c]{2}{*}{$0.00876$} & $St$
 & $0.785$ & $0.816$ & $0.863$ & $0.881$ & $0.767$
 & $0.485$ & $0.270$ & $0.142$ & $0.0297$ \\
 & & $Pe\,(\times 10^{-4})$
 & $0.270$ & $0.535$ & $1.04$ & $1.82$ & $2.08$
 & $1.56$ & $0.924$ & $0.501$ & $0.106$ \\
\hline

\multirow[c]{2}{*}{$15$} & \multirow[c]{2}{*}{$0.0296$} & $St$
 & $2.65$ & $2.75$ & $2.91$ & $2.97$ & $2.59$
 & $1.64$ & $0.913$ & $0.481$ & $0.100$ \\
 & & $Pe\,(\times 10^{-5})$
 & $0.136$ & $0.271$ & $0.525$ & $0.919$ & $1.05$
 & $0.788$ & $0.468$ & $0.253$ & $0.0539$ \\
\hline

\multirow[c]{2}{*}{$20$} & \multirow[c]{2}{*}{$0.0701$} & $St$
 & $6.28$ & $6.53$ & $6.90$ & $7.05$ & $6.14$
 & $3.88$ & $2.16$ & $1.14$ & $0.237$ \\
 & & $Pe\,(\times 10^{-5})$
 & $0.431$ & $0.856$ & $1.66$ & $2.91$ & $3.32$
 & $2.49$ & $1.48$ & $0.801$ & $0.170$ \\
\hline

\multirow[c]{2}{*}{$25$} & \multirow[c]{2}{*}{$0.137$} & $St$
 & $12.3$ & $12.7$ & $13.5$ & $13.8$ & $12.0$
 & $7.58$ & $4.23$ & $2.23$ & $0.464$ \\
 & & $Pe\,(\times 10^{-5})$
 & $1.05$ & $2.09$ & $4.05$ & $7.09$ & $8.11$
 & $6.08$ & $3.61$ & $1.96$ & $0.416$ \\
\hline

\multirow[c]{2}{*}{$30$} & \multirow[c]{2}{*}{$0.237$} & $St$
 & $21.2$ & $22.0$ & $23.3$ & $23.8$ & $20.7$
 & $13.1$ & $7.30$ & $3.85$ & $0.801$ \\
 & & $Pe\,(\times 10^{-6})$
 & $0.218$ & $0.433$ & $0.840$ & $1.47$ & $1.68$
 & $1.26$ & $0.749$ & $0.405$ & $0.0862$ \\
\hline

\multirow[c]{2}{*}{$40$} & \multirow[c]{2}{*}{$0.561$} & $St$
 & $50.2$ & $52.2$ & $55.2$ & $56.4$ & $49.1$
 & $31.1$ & $17.3$ & $9.12$ & $1.90$ \\
 & & $Pe\,(\times 10^{-6})$
 & $0.690$ & $1.37$ & $2.66$ & $4.65$ & $5.31$
 & $3.98$ & $2.37$ & $1.28$ & $0.273$ \\
\hline

\multirow[c]{2}{*}{$50$} & \multirow[c]{2}{*}{$1.10$} & $St$
 & $98.1$ & $102$ & $108$ & $110$ & $95.9$
 & $60.7$ & $33.8$ & $17.8$ & $3.71$ \\
 & & $Pe\,(\times 10^{-7})$
 & $0.168$ & $0.334$ & $0.649$ & $1.13$ & $1.30$
 & $0.973$ & $0.578$ & $0.313$ & $0.0665$ \\
\hline
\hline
\end{tabular}
\end{table*}

An important assumption we make here is that cloud droplets behave like perfect conductors, implying that electrical charges reside entirely on their surfaces. However, it is important to acknowledge that cloud droplets contain intrinsic dissolved ions resulting from their formation on soluble CCN, which bring salts and acids into the droplet solution upon activation. The ionic strength $I$ in cloud droplets can vary considerably depending on the aerosol source, liquid water content (LWC), and prevailing local chemical conditions, with reported values ranging from $10^{-5}$ to $10^{-2}$ mol L$^{-1}$ (see the review by \citet{herrmann2015tropospheric}). The Debye length $\lambda_D$, which quantifies the distance over which mobile ions effectively screen electrostatic interactions in an ionic solution, is given by $\lambda_D=[k_w \epsilon_0 k_B T/(2 N_A e^2 I)]^{1/2}$. Here, $k_w$ is the dielectric constant of the solvent (approximately 80 for water), and $N_A = 6.023 \times 10^{23}$ mol$^{-1}$ is Avogadro's number. Even for a cloud droplet with $I=10^{-5}$ mol L$^{-1}$, $\lambda_D$ is approximately $0.093$ \textmu m, a value that is substantially smaller than the typical radii of cloud droplets. In this scenario, an external electric field can polarize the droplet almost instantly, and dissolved ions rapidly migrate to the droplet's surface, effectively screening the interior. This is the origin of the assumption that cloud droplets act as perfect conductors.

Finally, we assume that the net charge on each droplet remains constant throughout the duration of droplet encounters. However, in reality, the conductivity of the surrounding air, due to the mobility of free ions, may result in a decrease in droplet charge. In this context, the instantaneous surface charge of a droplet with an initial charge $q_0$ can be expressed as $q_0\exp(-t/\tau)$, where $\tau=\epsilon_0/\sigma$ denotes the charge relaxation time, with $\sigma$ representing the conductivity of the air. For clear air at sea level, $\sigma \approx 2.3 \times 10^{-14}$ Sm$^{-1}$, which gives $\tau \approx 6.5$ min. However, the conductivity can vary significantly in cloud environments, ranging from $1/40$ to $1/3$ of the aforementioned value, primarily due to a decreased concentration of gas-phase free ions  \citep{pruppacher1997microphysics}. Based on these conductivity values, $\tau$ in clouds can vary approximately from $20$ min to $2$ h \citep{khain2004rain}. In contrast, the time scale for droplet-droplet encounters is approximately equivalent to the time scale of the problem $t^*$, which is about $4 \times 10^{-4}$ s for droplets of sizes $a_1 = 20$ \textmu m and $a_2 = 10$ \textmu m. Given that $t^* \ll \tau$, neglecting the variation in droplet surface charge during the collision events is justifiable.

In light of these assumptions, we will now present a comprehensive framework for calculating the relative trajectories and collision rates of two charged droplets settling in a quiescent fluid, while subjected to a vertical electric field.

\subsection{Equations of motion}\label{subsec:Equations of motion} 

The equations governing the motion of two spherical droplets, denoted as $i=1,2$, are given as follows:
\begin{subeqnarray}
    m_i\frac{\mathrm{d} \boldsymbol{V}_i}{\mathrm{d} t} & = & m_i \boldsymbol{g} + \boldsymbol{F}_i^H + \boldsymbol{F}_i^E + \boldsymbol{F}_i^{vdW}, \label{acceleration_equation} \\[3pt]
    \frac{\mathrm{d} \boldsymbol{X}_i}{\mathrm{d} t} & = & \boldsymbol{V}_i. \label{velocity_equation}
\end{subeqnarray}
In these equations, $\boldsymbol{V}_i$ and $\boldsymbol{X}_i$ represent the velocity and position of droplet $i$, respectively. The mass of droplet $i$ is defined by the relation $m_i=(4\pi/3)\rho_d a_i^3$. On the right-hand side of Eq. (\ref{acceleration_equation}a), $m_i\boldsymbol{g}$, $\boldsymbol{F}_i^H$, $\boldsymbol{F}_i^E$, and $\boldsymbol{F}_i^{vdW}$ represent the gravitational, hydrodynamic, electrostatic, and van der Waals forces, respectively. We define the relative velocity between the two droplets as $\boldsymbol{V} = \boldsymbol{V}_{1} - \boldsymbol{V}_{2}$, and the center-to-center separation vector pointing from the center of droplet $2$ to the center of droplet $1$ as $\boldsymbol{r}=\boldsymbol{X}_1-\boldsymbol{X}_2$.

\subsection{Hydrodynamic forces}\label{subsec:Hydrodynamic forces} 

The sedimentation of droplets within a fluid medium induces perturbations in the fluid around them, thereby generating a disturbance flow field. When the Reynolds number of the fluid flow around the settling droplets is small, the Stokes equations provide an accurate mathematical framework for describing the resultant flow field. Within the scope of microhydrodynamics, the resistance problem describes the relationship between the hydrodynamic forces and torques exerted on two spheres and their translational and angular velocities \citep{kim2013microhydrodynamics}. Notably, the forces acting upon fluid droplets are independent of their angular velocities, distinguishing them from the hydrodynamic force expressions applicable to rigid spherical particles. This noteworthy distinction arises because droplets experience no hydrodynamic torque under the Stokes flow conditions \citep[e.g.,][]{zinchenko1980slow}. In consequence, the hydrodynamic forces exerted on two droplets settling in a quiescent fluid can be expressed through linear relationships that depend solely on their translational velocities, as given by the following equations \citep{pigeonneau2002collision}:
\begin{subeqnarray}
    \boldsymbol{F}_1^H & = & -6\pi\mu_fa_1\left(\boldsymbol{A}_{11}\boldsymbol{\cdot}\boldsymbol{V}_1 + \boldsymbol{A}_{12}\boldsymbol{\cdot}\boldsymbol{V}_2\right), \label{hydrodynamic_force_on_droplet_1} \\[3pt]
    \boldsymbol{F}_2^H & = & -6\pi\mu_fa_2\left(\boldsymbol{A}_{21}\boldsymbol{\cdot}\boldsymbol{V}_1 + \boldsymbol{A}_{22}\boldsymbol{\cdot}\boldsymbol{V}_2\right). \label{hydrodynamic_force_on_droplet_2}
\end{subeqnarray}
Here, $\boldsymbol{A}_{\alpha \beta}$ (where $\alpha$ and $\beta$ are either $1$ or $2$) represent second-rank resistance tensors defined as: 
\begin{equation}
    \boldsymbol{A}_{\alpha \beta}=X_{\alpha\beta}^A \frac{\boldsymbol{r} \boldsymbol{r}}{r^2} + Y_{\alpha\beta}^A\left(\boldsymbol{I}-\frac{\boldsymbol{r} \boldsymbol{r}}{r^2}\right),
\end{equation}
where $r=|\boldsymbol{r}|$, and $\boldsymbol{I}$ is the identity tensor of second order. The dimensionless scalar quantities $X_{\alpha\beta}^A$ and $Y_{\alpha\beta}^A$ are axisymmetric and asymmetric resistance functions, respectively, corresponding to 
droplet relative motion along their line of centers and normal to that line, respectively. These resistance functions depend on $\xi$, $\kappa$, and $\hat{\mu}$.

Various methods exist in the literature for computing these resistance functions by solving the Stokes equations for a system comprising two hydrodynamically interacting spheres. \cite{jeffrey1984calculation} employed twin multipole expansions to derive the resistance functions for two rigid spheres. Here we utilize the results of \cite{haber1973low,zinchenko1982calculation} and \cite{zinchenko1980slow}, who developed exact solutions of the Stokes equations in bispherical coordinates to calculate axisymmetric and asymmetric resistance functions for two spherical droplets. Using the method of reflections, \cite{hetsroni1978low} derived the far field (i.e., $\xi \gg 1$) analytical expressions of these functions as series in inverse powers of $\xi$. In the lubrication regime ($\xi \ll 1$), \cite{davis1989lubrication} derived asymptotic expressions for axisymmetric resistance functions employing a boundary collocation technique. They provided the expressions for $X_{\alpha\beta}^A$ in terms of Pad\'e approximants as
\begin{subeqnarray}
    X_{11}^A(\xi,\kappa,\hat{\mu}) & = & \kappa X_{22}^A(\xi,\kappa,\hat{\mu})= \frac{2\kappa^2}{(1+\kappa)^3\xi}\mathcal{R}(\eta) \text{~~and} \label{lubrication_forms_X11A_X22A} \\[3pt]
     X_{12}^A(\xi,\kappa,\hat{\mu}) & = & \kappa X_{21}^A(\xi,\kappa,\hat{\mu})=-X_{11}^A(\xi,\kappa,\hat{\mu}), \label{lubrication_forms_X12A_X21A}
\end{subeqnarray}
where
\refstepcounter{equation}
$$
  \mathcal{R}(\eta) = \frac{1+0.38 \eta}{1+1.69 \eta + 0.43 \eta^2} \quad\text{and}\qquad
  \eta = \frac{1}{\hat{\mu}}\left(\frac{2\kappa}{(1+\kappa)^2\xi}\right)^{1/2}.
  \eqno{(\theequation{\mathit{a},\mathit{b}})} \label{Expresions_for_f(m)_and_m}
$$
The function $\mathcal{R}(\eta)$ characterizes the ratio of axisymmetric resistance functions for spherical droplets compared to those of rigid spherical particles. For water droplets dispersed in air, with a viscosity ratio $\hat{\mu}=10^2$, the value of $\mathcal{R}(\eta)$ can be as low as approximately $0.526$ when $\xi=10^{-4}$. Consequently, despite the significant air--water viscosity contrast, when examining the collision dynamics of cloud droplets it is essential to account for the internal flows within the droplets. For $Y_{\alpha\beta}^A$, no lubrication solutions exist. Importantly, these asymmetric resistance functions do not exhibit singular behavior as the separation distance between the two droplets approaches zero.  Evidently, the only relevant estimate is that of \cite{zinchenko1982calculation}, which is 
\begin{equation}
    Y_{\alpha\beta}^A(\xi,\kappa,\hat{\mu}) = Y_{\alpha\beta}^A(0,\kappa,\hat{\mu}) + O(|\ln\xi|^{-1}). \label{lubrication_forms_YA}
\end{equation}

\subsection{Electrostatic forces}\label{subsec:Electrostatic forces}

The determination of electrostatic interactions between two conducting spheres is fundamentally a boundary value problem. Therefore, it is necessary to solve Laplace's equation for the potential field in the region outside the two spheres. The electrostatic forces acting on each sphere can be derived by integrating the electrical stresses over their respective surfaces. As noted above, \citet{davis1964two} provided a comprehensive analysis of the electrostatic forces experienced by two charged spherical conductors placed in a uniform external electric field by solving for the potential field in a bispherical coordinate system. These forces depend on various factors, including the relative geometry of the two spheres, the magnitude and polarity of surface charges on each sphere, the intensity of the external electric field, $E_0$, and the angle $\theta$ formed between the direction of the electric field $\boldsymbol{E}_0$ and the line joining the centers of the two spheres. The axisymmetric nature of the two-sphere configuration allows for the decomposition of the force acting on each sphere into two distinct components: one parallel to the line connecting the centers of the two spheres and the other perpendicular to this line. The expression for the electrostatic force exerted on the droplet of radius $a_2$ is given as
\begin{eqnarray}
    \boldsymbol{F}_2^{E} = \bigg[4\pi \epsilon_0 a_2^2 E_0^2\left(F_1\cos^2\theta+F_2\sin^2\theta\right)+E_0\cos\theta\left(F_3 q_1+F_4 q_2\right) \nonumber \\  + \frac{1}{4\pi \epsilon_0 a_2^2}\left(F_5 q_1^2+F_6 q_1 q_2+F_7 q_2^2\right)+E_0 q_2\cos\theta \bigg]\boldsymbol{\hat{e}}_r \nonumber \\ - \bigg[4\pi \epsilon_0 a_2^2 E_0^2 F_8\sin2\theta+E_0\sin\theta\left(F_9 q_1+F_{10}q_2\right)+E_0 q_2\sin\theta \bigg]\boldsymbol{\hat{e}}_{\theta}, \label{Electrostatic_force_on_droplet_2}
\end{eqnarray}
where $q_1$ and $q_2$ are the charges on droplets with radius $a_1$ and $a_2$, respectively, $F_1, F_2, \ldots, F_{10}$ are the dimensionless force coefficients that depend on $\xi$ and $\kappa$. In appendix \ref{appA}, we  discuss these force coefficients and give analytical expressions for some of them in both the far and near fields.

The quantity $\boldsymbol{E}_0(q_1+q_2)$ is the external electric force exerted on a two sphere system. Because the vector sum of the electrostatic forces on the two spheres, $\boldsymbol{F}_1^{E} + \boldsymbol{F}_2^{E}$, must equal the external electric force acting on the system, then the electrostatic force acting on the droplet of radius $a_1$ is
\begin{equation}
  \boldsymbol{F}^{E}_1 = \boldsymbol{E}_0\left(q_1+q_2\right)-\boldsymbol{F}^{E}_2. \label{Electrostatic_force_on_droplet_1}
\end{equation}

\subsection{van der Waals forces}\label{subsec:van der Waals forces}

Intermolecular forces, particularly the attractive van der Waals (vdW) force between like materials, play a crucial role in the collisions between relatively small cloud droplets. The van der Waals force always acts along the line connecting the centers of the two droplets, allowing us to express the forces on the droplets as $\boldsymbol{F}_1^{\text{vdW}}=-\boldsymbol{F}_2^{\text{vdW}}=-(\mathrm{d} \varPhi_{\text{vdW}}/\mathrm{d} r) \boldsymbol{\hat{e}}_r$, where $\mathrm{d} \varPhi_{\text{vdW}}/\mathrm{d} r$ represents the gradient of the van der Waals potential $\varPhi_{\text{vdW}}$. \citet{hamaker1937london} derived an analytical expression for the non-retarded vdW potential using a theory based on pairwise additivity. However, when the separation distance between atoms is large, retardation effects arise due to the frequency dependence of the finite speed of electromagnetic wave propagation (see \citet{RevModPhys.82.1887}, for a review), leading to a weakening of the van der Waals force \citep{casimir1948influence}. Effects of retardation must be considered when the droplet separation distance is comparable to or exceeds the London wavelength $\lambda_L (\approx 0.1$ \textmu m). Numerous expressions have been proposed to describe retarded van der Waals forces, and they typically yield consistent results. The retarded van der Waals forces depend on several parameters, including the Hamaker constant $A_H$, the nondimensional quantity $N_L = 2 \pi \left(a_1+a_2\right)/\lambda_L = 2 \pi a_1 \left(1+\kappa\right)/\lambda_L)$, $\xi$, and $\kappa$ (see \citet{zinchenko1994gravity}). The value of $A_H$ for water droplets in air is approximately $3.7 \times 10^{-20}$ J \citep{friendlander2000smoke}.

\subsection{Expressions for the collision rate and efficiency}\label{subsec:Expressions_for_the_collision_rate_and_efficiency}

The collision rate, denoted as $K_{12}$, is given by the following expression,
\begin{equation}
    K_{12} = -n_1n_2\int_{\mathcal{S}}\left(\boldsymbol{V}\boldsymbol{\cdot}\boldsymbol{n}\right) P \hspace{0.5mm} \mathrm{d}A. \label{Collision_rate_general_dimensional}
\end{equation}
Here, $\mathcal{S}$ represents the region on the collision surface defined by $r=a_1+a_2$, where the condition $\boldsymbol{V}\boldsymbol{\cdot}\boldsymbol{n} < 0$ holds, and $\boldsymbol{n}$ is the outward unit normal vector to the collision surface. A key challenge in this formulation is accurately determining the pair-distribution function $P(r)$. For dilute systems, such as atmospheric clouds, the pair-distribution function is governed by the quasi-steady Fokker-Planck equation:
\begin{equation}
  \boldsymbol{\nabla}\boldsymbol{\cdot}\left(P\boldsymbol{V}\right) = 0.
  \label{Probability_conservation}
\end{equation}
The uncorrelated motion of the droplets in the far field defines the boundary condition: $P \rightarrow 1$ as $r \rightarrow \infty$. For computational purposes, we consider $r\equiv r_{\infty}$, which is defined as significantly large yet finite.

Using equation (\ref{Probability_conservation}) with the divergence theorem, the collision rate becomes
\begin{equation}
    K_{12} = -n_1n_2\int_{\mathcal{S'}}\left.\left(\boldsymbol{V}\boldsymbol{\cdot}\boldsymbol{n'}\right)\right\vert_{r=r_{\infty}} P \hspace{0.5mm} \mathrm{d}A,
    \label{Collision_rate_upstream_interception_area_1}
\end{equation}
where $\mathcal{S'}$ denotes the upstream interception area and $\boldsymbol{n'}$ is the unit outward normal vector to the area element of $\mathcal{S'}$. Specifically, $\mathcal{S'}$ represents the upstream cross-sectional area of the volume occupied by all trajectories originating from $r=r_{\infty}$ and terminating at $r=a_1+a_2$. At $r=r_{\infty}$, we have $P=1$ and $\boldsymbol{V}\cdot\boldsymbol{n'} = -V_{12}^{(0)}$, thereby simplifying equation (\ref{Collision_rate_upstream_interception_area_1}) to
\begin{equation}
    K_{12} = n_1n_2 V_{12}^{(0)} \pi y_c^2, \label{Collision_rate_upstream_interception_area_2}
\end{equation}
where $y_{c}$ is the critical impact parameter, quantifying the maximum horizontal distance between the centers of two widely separated droplets that can still collide. In this case, the relative trajectory formed by two colliding droplets is called the limiting colliding trajectory. These limiting colliding trajectories define the boundary that separates other colliding trajectories from non-colliding trajectories. In this problem, the upstream interception area takes the shape of a circle with radius $y_c$, as illustrated schematically in figure \ref{Schematic_binary_collisions_charged_droplets_with_gravity_electric_field}$b$. The collision rate in the absence of any interactions is described by the classical Smoluchowski model \citep{smoluchowski1918versuch}, where $y_{c}=a_1+a_2$ (see figure \ref{Schematic_binary_collisions_charged_droplets_with_gravity_electric_field}$b$). Therefore, the expression for the ideal collision rate, denoted as $K_{12}^{0}$, is given by
\begin{equation}
K_{12}^{0} = n_1n_2V^{(0)}_{12}\pi \left(a_1+a_2\right)^2.
\label{ideal_collision_rate}
\end{equation}
Finally, the collision efficiency, denoted by $E_{12}$, is defined as the ratio of $K_{12}$ to $K_{12}^{0}$:
\begin{equation}
E_{12} = \frac{K_{12}}{K_{12}^{0}} = \frac{y_{c}^2}{\left(a_1+a_2\right)^2}. 
\label{Collision_efficiency}
\end{equation}

To compute the collision rate and efficiency, it is necessary to identify the maximum horizontal shift beyond the region of interdroplet interaction, where collisions between droplets may still occur. In section \ref{sec:num} below, we outline the numerical method for determining $y_c$ by analyzing the relative trajectories of a pair of droplets.

\subsection{Numerical method\label{sec:num}}

We numerically integrate the governing equations (\ref{acceleration_equation}) to track the trajectories of two inertial droplets, considering both hydrodynamic and non-hydrodynamic interactions. As the droplets approach each other, the interaction forces increase significantly, resulting in a stiff system of ordinary differential equations. To effectively address this complexity, we employ a fourth-order Runge-Kutta algorithm with an adaptive time step, thereby ensuring both the accurate resolution of these equations and the physical consistency of the simulations by preventing droplet overlap. This algorithm is quite similar to that presented in \citet{dubey2024critical}.

Since the electrostatic and van der Waals forces acting on the droplets do not depend on the azimuthal coordinate, the relative motion of the droplets remains axisymmetric about the vertical axis. Therefore, it is sufficient to analyze the dynamics in the $y$-$z$ plane while constraining the $x$-component of velocity for each droplet to zero across all time steps. Initially, we place the center of the smaller droplet at the origin $(0, 0, 0)$, while the centers of several larger droplets are positioned at $(0, y_0, z_0)$, treating each configuration as an independent problem. We set the value of $z_0$ to $100(a_1 + a_2)$, which is sufficiently large to ensure that the collision dynamics are independent of the initial center-to-center distance in the vertical direction. For the initial horizontal shift, we consider different values within the entire possible range of collision, denoted as $[0, L_h]$. The upper limit, $L_h$, in this range varies in our simulation, depending on the sizes of the droplets, their electric charges, and the intensity of the external electric field. At large initial separation distances, the motion of the droplets becomes nearly independent of one another. Consequently, we use the terminal velocities of the droplets settling independently in quiescent air as the initial conditions in the momentum equations (\ref{acceleration_equation}$a$). It is essential to note that two widely separated charged droplets in a vertical electric field will experience additional electrostatic forces, whose magnitudes and directions depend on the amount and polarity of charges on the droplets and the strength of the electric field. Therefore, we take into account these electrostatic forces acting on the droplets while calculating their terminal velocities.

According to the definition in (\ref{Collision_rate_general_dimensional}), the radial relative velocity between two colliding droplets must remain negative until the moment of contact. To identify collision events in our simulations, we assume that the droplets touch each other when the non-dimensional separation distance between them reaches $\xi=10^{-5}$. We have confirmed that using a smaller cutoff for the separation distance does not alter the value of the collision efficiency. Starting with $y_0=L_h$, we methodically evolve droplet pairs sequentially for a series of consecutive $y_0$ values, each spaced decrementally by $10^{-2}a^*$. We record the $y_0$ value and stop the integration upon detecting the first collision event. This recorded $y_0$ value corresponds to $y_c$. 

\section{Relative trajectory and collision efficiency} \label{sec:Relative_trajectory_and_collision_efficiency}

Our primary objective is to quantitatively evaluate the influence of electrostatic interactions on the collision dynamics of two spherical water droplets settling in still air, using the trajectory analysis methodology outlined in the preceding section. To establish a clear context, we first briefly review the relative trajectories of two hydrodynamically interacting sedimenting droplets without electrostatic interactions, that is when $q_1=q_2=0$ and $E_0=0$. We then present typical pair trajectories and assess collision efficiency by incorporating the effects of electrostatic forces on droplet motion. As previously mentioned, the problem exhibits axisymmetry, with the vertical axis serving as the axis of symmetry. Thus, we can analyze the relative trajectories of a droplet pair in a representative $r\sin\theta - r\cos\theta$ plane. We non-dimensionalize spatial coordinates by $a^*$ while presenting the relative trajectories. In all simulations, we use the specified values for the densities of water droplets $\rho_d$ and air $\rho_f$, the dynamic viscosity of air $\mu_f$, the droplet-to-medium viscosity ratio $\hat{\mu}$, and the Hamaker constant $A_H$, as stated in \S \ref{sec:Introduction}.

\begin{figure}
\centering
\includegraphics[width=1.0\textwidth]{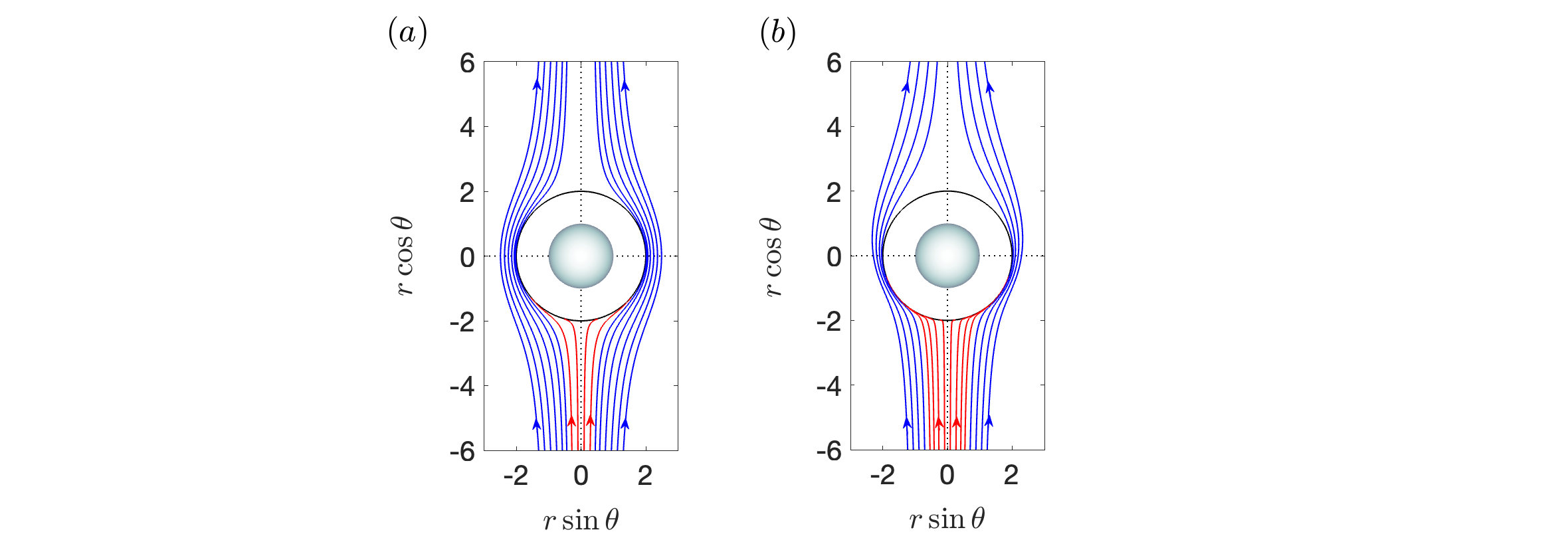}
\caption{The trajectories of a droplet with a radius of $20$ \textmu m relative to a droplet with a radius of $10$ \textmu m without considering electrostatic interactions when $(a)$ $St=0$ and $(b)$ $St=6.87$, where $St = \tau_d/t^* = 2C_v\rho_d(\rho_d-\rho_f) g a_1^3(1+\kappa)(1-\kappa^2)/(81\mu_f^2)$. Coordinates in these plots are scaled by $a^*$. In both scenarios, initial conditions for the corresponding trajectories are identical. The sphere at the center represents the test droplet, and the thin black circle represents the projection of the collision sphere. Arrows along the trajectories indicate their direction.}
\label{Relative_trajectories_for_a1_20_a2_10_with_vdW_no_electrostatics_with_and_without_inertia}
\end{figure}

To demonstrate the influence of droplet inertia, we examine the relative trajectories of a pair of droplets with radii $a_1=20$ \textmu m and $a_2=10$ \textmu m, subject solely to gravitational forces. We consider two distinct scenarios: (a) droplets treated as inertia-free and (b) droplet inertia taken into account (see figure \ref{Relative_trajectories_for_a1_20_a2_10_with_vdW_no_electrostatics_with_and_without_inertia}). In both scenarios, we identify two types of relative trajectories: open (shown by blue lines) and colliding (shown by red lines). We use an identical set of initial horizontal offsets between the two droplets in both cases to compare their relative trajectory maps. The comparison between the two scenarios suggests that droplet inertia can lead to collisions even when initial horizontal offsets between the droplets are relatively larger. Furthermore, for $St=0$, open trajectories are symmetric about the $r \cos \theta = 0$ axis. However, this symmetry is broken in the presence of droplet inertia. For low Stokes numbers, specifically when $St < O(10^{-1})$, hydrodynamic forces dominate, and the trajectories closely resemble those at $St=0$. As the Stokes number increases, the influence of hydrodynamic effects diminishes, leading to ballistic collisions of the droplets at higher $St$.

Van der Waals forces introduce a non-linearity into the governing equations, even in the absence of droplet inertia. This non-linearity results in asymmetry in the relative trajectories, similar to what we observed for inertial droplets (see figure \ref{Relative_trajectories_for_a1_20_a2_10_with_vdW_no_electrostatics_with_and_without_inertia}$b$). We define the types of collision between two droplets based on their impact locations on the collision sphere as follows: if a colliding trajectory hits the collision sphere where the angle $\theta$ falls within the range $[\pi/2, 3\pi/2]$, we refer to this as a front-side collision between the droplets, otherwise, if a colliding trajectory hits the collision sphere outside this angle range, we classify it as a rear-side collision. As the droplets approach each other closely, the attractive van der Waals forces can pull them together, resulting in rear-side collisions. However, when droplet inertia is significant, the influence of van der Waals forces becomes negligible; consequently, whether these attractive van der Waals forces are accounted for or disregarded results in nearly indistinguishable relative trajectories. Although we have not explicitly shown the relative trajectories with van der Waals forces, examples are given in figure 4 of \citealt{rother2022gravitational}.

Electrostatic interactions that affect cloud droplet collision dynamics depend on several factors, including droplet sizes, the magnitude and polarity of droplet surface charge, and the strength of the electric field in clouds. Here we focus on droplets that are either electrically neutral or possess surface charges with magnitudes given by the formula: $|q_i| = 2 \times 3.33 \times 10^{-6} a_i^2$ ($i=1,2$), where $q_i$ is the charge (in coulombs) on droplet $i$, with a corresponding radius $a_i$ (in meters). Thus, the charge ratio $\kappa_q$, defined as $\kappa_q=q_2/q_1$, depends solely on the ratio of the droplet radii and can be either $+\kappa^2$ or $-\kappa^2$. The electric field strengths we consider are $0$, $5 \times 10^4$, and $3 \times 10^5$ Vm$^{-1}$. Table \ref{Different_cases_of_electric_charg_and_field_of_droplets} presents various combinations of the surface charge and electric field strengths that we consider. To compare the relative trajectories across all of these cases, we consider $a_1=20$ \textmu m and $a_2=10, 18$ \textmu m, while presenting their corresponding relative trajectories.

\begin{table*}
\centering
\caption{Droplet charges $q_1$ and $q_2$ and the vertical electric field strength $E_0$ for the eight cases considered in this study. The droplet radii $a_1$ and $a_2$ are in meters.}
\label{Different_cases_of_electric_charg_and_field_of_droplets}
\renewcommand{\arraystretch}{1.15}
\begin{tabular}{*{4}{c}}
\hline
\hline
Case & $q_1$ ($\mathrm{C}$) & $q_2$ ($\mathrm{C}$)
     & $E_0$ ($\mathrm{V\,m^{-1}}$) \\
\hline
I    & $2 \times 3.33 \times 10^{-6} a_1^2$
     & $2 \times 3.33 \times 10^{-6} a_2^2$  & $0$ \\
II   & $2 \times 3.33 \times 10^{-6} a_1^2$
     & $-2 \times 3.33 \times 10^{-6} a_2^2$ & $0$ \\
III  & $0$ & $0$ & $5 \times 10^4$ \\
IV   & $0$ & $0$ & $3 \times 10^5$ \\
V    & $2 \times 3.33 \times 10^{-6} a_1^2$
     & $2 \times 3.33 \times 10^{-6} a_2^2$  & $5 \times 10^4$ \\
VI   & $2 \times 3.33 \times 10^{-6} a_1^2$
     & $2 \times 3.33 \times 10^{-6} a_2^2$  & $3 \times 10^5$ \\
VII  & $2 \times 3.33 \times 10^{-6} a_1^2$
     & $-2 \times 3.33 \times 10^{-6} a_2^2$ & $5 \times 10^4$ \\
VIII & $2 \times 3.33 \times 10^{-6} a_1^2$
     & $-2 \times 3.33 \times 10^{-6} a_2^2$ & $3 \times 10^5$ \\
\hline
\hline
\end{tabular}
\end{table*}

\begin{figure}
\centering
\includegraphics[width=1.0\textwidth]{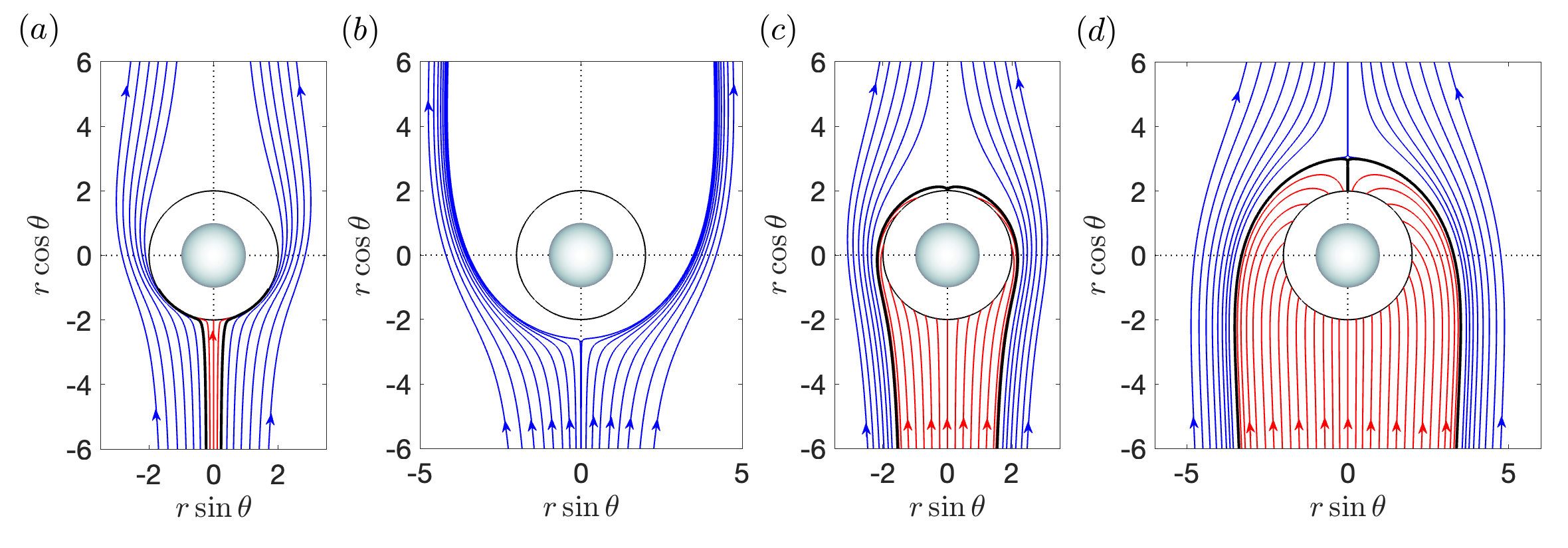}
\caption{Relative trajectories of two differentially settling charged cloud droplets without any external electric field for $(a)$ case I with $a_1=20$ \textmu m and $a_1=10$ \textmu m, $(b)$ case I with $a_1=20$ \textmu m and $a_1=18$ \textmu m, $(c)$ case II with $a_1=20$ \textmu m and $a_2=10$ \textmu m and $(d)$ case II with $a_1=20$ \textmu m and $a_2=18$ \textmu m. The color scheme for the trajectory types is consistent with that used in figures \ref{Schematic_binary_collisions_charged_droplets_with_gravity_electric_field}$(b)$ and \ref{Relative_trajectories_for_a1_20_a2_10_with_vdW_no_electrostatics_with_and_without_inertia}.}
\label{Relative_trajectories_for_cases_I_and_II}
\end{figure}

The collision dynamics of charged sedimenting droplets, in the absence of an external electric field, depend significantly on the magnitude and polarity of their electrical charges. For weakly charged droplets, the collision dynamics primarily depend on hydrodynamic forces and resemble those of uncharged droplets. However, as the magnitude of the charge increases, the influence of hydrodynamic forces diminishes, resulting in the dynamics increasingly being controlled by electrostatic forces. Figure \ref{Relative_trajectories_for_cases_I_and_II}$a$ shows the relative trajectories of two differentially sedimenting droplets with radii $a_1=20$ \textmu m and $a_2=10$ \textmu m, both carrying charges of the same polarity, as described in case I of Table \ref{Different_cases_of_electric_charg_and_field_of_droplets}. Despite the droplets carrying charges of the same polarity, the electrostatic force can be attractive at small separation distances due to the extreme polarization of their charges, ultimately leading to collisions between them. The fixed points of the relative droplet dynamics are determined from $\boldsymbol{\dot{r}}=0$ and $\boldsymbol{\dot{V}}=0$, one of which was identified as a saddle point located on the $r\sin\theta=0$ axis above the collision sphere (see \citealt{dubey2024critical}). The limiting colliding trajectories, shown as thick black lines, act as separatrices that tend to converge at the saddle point, which resides close to the collision sphere. In other words, the stable manifold of the saddle point forms the separatrices. As in \citet{dubey2024critical}, we were unable to resolve the saddle point in our numerical simulations, as the numerical integration becomes challenging when the trajectory travels along the collision sphere at small interdroplet separation distances, where lubrication interactions dominate. When the electrostatic force between like-charged droplets dominates over gravitational and hydrodynamic forces at large separation distances, the Coulomb repulsion prevents droplets from coming into contact, resulting in only non-colliding relative trajectories (see \ref{Relative_trajectories_for_cases_I_and_II}$b$). The saddle point in this case results from the balance between differential settling and electrostatic repulsion forces. Evidently, the saddle point appears below the collision sphere, and its location on the $r\sin\theta=0$ axis is far from the collision sphere. Unlike the previous case, the stable manifold of the saddle point in this case is one-dimensional and thus cannot form separatrices.

Figure \ref{Relative_trajectories_for_cases_I_and_II}$c$-$d$ illustrates the relative trajectories of droplet pairs with opposite charges, corresponding to case II in Table \ref{Different_cases_of_electric_charg_and_field_of_droplets}. The sizes of the droplet pairs in these two scenarios are the same as those considered in figure \ref{Relative_trajectories_for_cases_I_and_II}$a$-$b$. The critical impact parameter in figure \ref{Relative_trajectories_for_cases_I_and_II}$c$ is significantly larger than that in figure \ref{Relative_trajectories_for_cases_I_and_II}$a$. This difference arises because the electrostatic forces acting on oppositely charged droplets are always attractive at large separation distances. Furthermore, at small separation distances, the magnitude of the attractive force between droplets with opposite charges is stronger than that between droplets with charges of the same polarity. In figure \ref{Relative_trajectories_for_cases_I_and_II}$c$, the saddle point lies slightly away from the collision surface, making it clear that the separatrices (the trajectories that limit collisions) approach the collision sphere at nearly $\theta=0$. This type of relative trajectory dynamics is even more pronounced in figure \ref{Relative_trajectories_for_cases_I_and_II}$d$, where attractive electrostatic forces dominate the collision dynamics. Here, since the saddle point and its stable manifold are far from the collision sphere, the near-field hydrodynamics do not significantly influence the collision dynamics.

\begin{figure}
\centering
\includegraphics[width=1.0\textwidth]{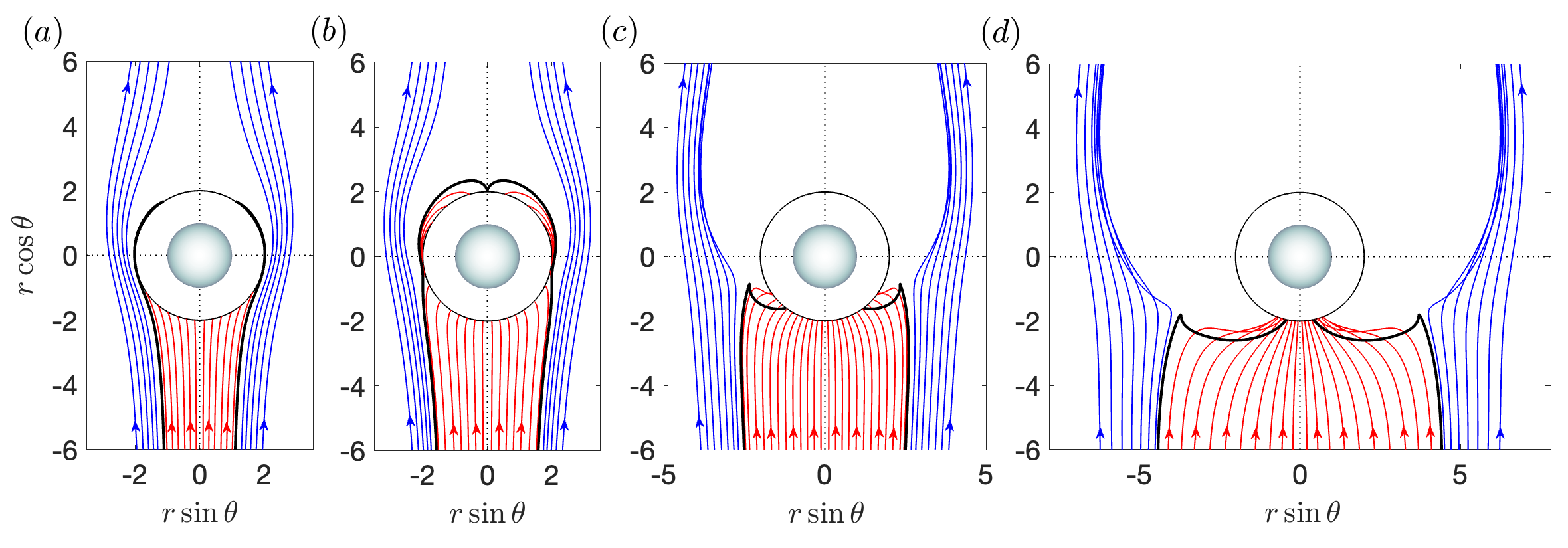}
\caption{Relative trajectories of two uncharged cloud droplets subject to gravitational force and a vertical electric field for $(a)$ case III with $a_1=20$ \textmu m and $a_1=10$ \textmu m, $(b)$ case III with $a_1=20$ \textmu m and $a_1=18$ \textmu m, $(c)$ case IV with $a_1=20$ \textmu m and $a_2=10$ \textmu m and $(d)$ case IV with $a_1=20$ \textmu m and $a_2=18$ \textmu m. The color scheme for the trajectory types is the same as used in figure \ref{Relative_trajectories_for_cases_I_and_II}.}
\label{Relative_trajectories_for_cases_III_and_IV}
\end{figure}

To investigate the effects of an external vertical electric field on collision dynamics, we present typical relative trajectories of two uncharged settling droplets in figure \ref{Relative_trajectories_for_cases_III_and_IV}. The electric field strengths considered here are $5 \times 10^4$ Vm$^{-1}$ (case III) and $3 \times 10^5$ Vm$^{-1}$ (case IV). The sizes of the droplet pairs are consistent with those considered in figure \ref{Relative_trajectories_for_cases_I_and_II}. As anticipated, when the electric field-induced forces are relatively weak, the relative trajectories exhibit characteristics similar to those observed in the absence of electrostatic interactions (see figure \ref{Relative_trajectories_for_cases_III_and_IV}$a$). However, as the strength of the electric field-induced forces increases, the topology of the relative trajectories undergoes significant modification, even under moderate field effects (see figure \ref{Relative_trajectories_for_cases_III_and_IV}$b$). Under strong electric fields, the attractive forces facilitate collisions between droplets, even when the initial horizontal separation between their centers is considerably larger than in cases with weaker electric fields. The forces and torques induced by the strong electric field lead to pronounced deflection of relative trajectories at locations where the effects of electrostatic forces and differential settling are comparable (see figure \ref{Relative_trajectories_for_cases_III_and_IV}$c$-$d$). Importantly, all colliding trajectories tend to impact the collision sphere at locations centered around $\theta=\pi/2$, thereby precluding the occurrence of rear-side collisions. Due to the lack of appropriate initial conditions for the velocities of the individual droplets in the near field, we cannot present all possible relative trajectories that could be exhibited in the $St=0$ limit, as discussed by \citet{patra2025gravity}. Therefore, we are unable to numerically determine the locations of the fixed points and their nature of stability. Unlike in the $St=0$ scenario, relative trajectories can intersect each other, even if the trajectories of individual droplets in their respective phase spaces do not intersect.

\begin{figure}
\centering
\includegraphics[width=1.0\textwidth]{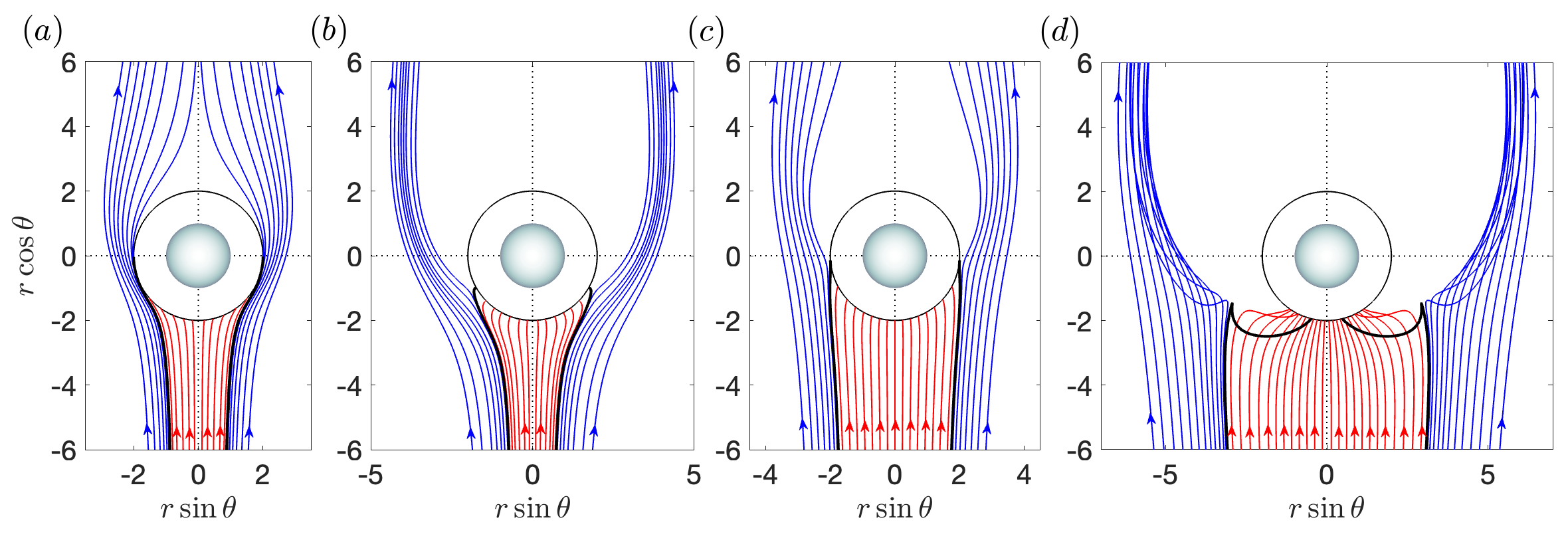}
\caption{Relative trajectories of two like-charged cloud droplets subject to gravitational force and a vertical electric field for $(a)$ case V with $a_1=20$ \textmu m and $a_1=10$ \textmu m, $(b)$ case V with $a_1=20$ \textmu m and $a_1=18$ \textmu m, $(c)$ case VI with $a_1=20$ \textmu m and $a_2=10$ \textmu m and $(d)$ case VI with $a_1=20$ \textmu m and $a_2=18$ \textmu m. The color scheme for the trajectory types is the same as used in figure \ref{Relative_trajectories_for_cases_I_and_II}.}
\label{Relative_trajectories_for_cases_V_and_VI}
\end{figure}

\begin{figure}
\centering
\includegraphics[width=1.0\textwidth]{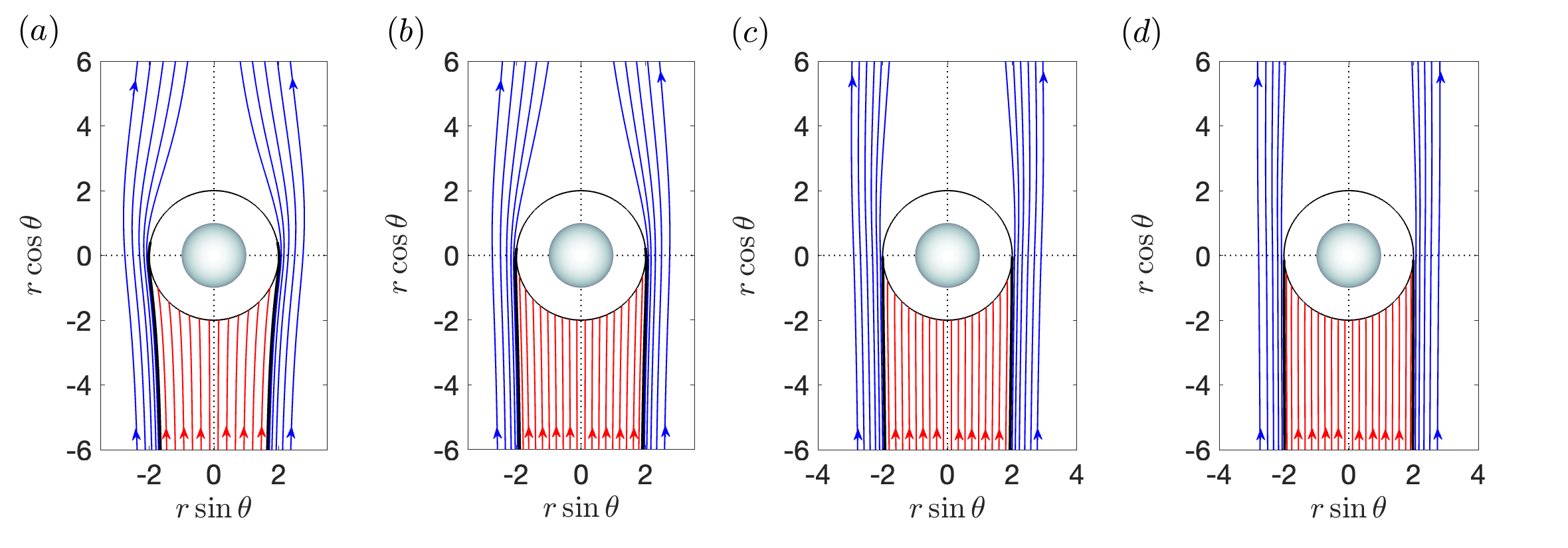}
\caption{Relative trajectories of two oppositely charged cloud droplets subject to gravitational force and a vertical electric field for $(a)$ case VII with $a_1=20$ \textmu m and $a_1=10$ \textmu m, $(b)$ case VII with $a_1=20$ \textmu m and $a_1=18$ \textmu m, $(c)$ case VIII with $a_1=20$ \textmu m and $a_2=10$ \textmu m and $(d)$ case VIII with $a_1=20$ \textmu m and $a_2=18$ \textmu m. The color scheme for the trajectory types is the same as used in figure \ref{Relative_trajectories_for_cases_I_and_II}.}
\label{Relative_trajectories_for_cases_VII_and_VIII}
\end{figure}

We will now illustrate the typical relative trajectories resulting from the combined effects of droplet surface charges and an external electric field. Figure \ref{Relative_trajectories_for_cases_V_and_VI} shows the representative relative trajectories for cases V and VI as outlined in Table \ref{Different_cases_of_electric_charg_and_field_of_droplets}. When comparing the relative trajectory maps for like-charged droplets under an electric field of strength $5 \times 10^4$ Vm$^{-1}$ (shown in figure \ref{Relative_trajectories_for_cases_V_and_VI}$a$-$b$) to those without an electric field (figure \ref{Relative_trajectories_for_cases_I_and_II}$a$-$b$), we observe that the collisional areas in the former cases are significantly larger. This increase in collisional area can be attributed to the electric field inducing opposite charges on the nearest surfaces of the droplets, thereby creating attractive forces between them. However, when we compare figures \ref{Relative_trajectories_for_cases_V_and_VI}$a$-$b$ with cases involving uncharged droplets subjected to the same electric field strength of $5 \times 10^4$ Vm$^{-1}$ (shown in figure \ref{Relative_trajectories_for_cases_III_and_IV}$a$-$b$), it is evident that the collisional areas in these cases are relatively small due to the repulsive effects of like charges. We observe similar trends with a higher electric field strength of $3 \times 10^5$ Vm$^{-1}$ when comparing figure \ref{Relative_trajectories_for_cases_V_and_VI}$c$-$d$ with figures \ref{Relative_trajectories_for_cases_I_and_II}$a$-$b$ and \ref{Relative_trajectories_for_cases_III_and_IV}$c$-$d$. As expected, the trajectory maps in figures \ref{Relative_trajectories_for_cases_V_and_VI}$c$-$d$ tend to resemble those in figures \ref{Relative_trajectories_for_cases_III_and_IV}$c$-$d$, which do not account for droplet surface charge effects.

Figure \ref{Relative_trajectories_for_cases_VII_and_VIII} presents examples of relative trajectories for oppositely charged droplets subjected to an external electric field, treating cases VII and VIII of Table \ref{Different_cases_of_electric_charg_and_field_of_droplets}. Unlike like-charged droplets, where the electric field consistently promotes collisions, the interaction of two oppositely charged droplets with an external electric field may either promote or inhibit collisions depending on a combination of factors, including the magnitudes of the charges, the sizes of the droplets, and the strength of the applied electric field. In this analysis, the magnitude of a droplet's net charge is related to its size. Thus, the impact of the electric field on the collision dynamics between oppositely charged droplets varies according to the sizes of the droplet pairs and the strength of the field. For example, in the case of two oppositely charged droplets with radii $a_1=20$ \textmu m and $a_1=10$ \textmu m, the upstream collisional area increases slightly when the externally applied electric field strengths are $5 \times 10^4$ Vm$^{-1}$, and $3 \times 10^5$ Vm$^{-1}$ (see figures \ref{Relative_trajectories_for_cases_VII_and_VIII}$a$,$c$ and \ref{Relative_trajectories_for_cases_I_and_II}$c$). Conversely, when considering droplets of opposite charges with radii $a_1=20$ \textmu m and $a_1=18$ \textmu m, the same electric field strengths lead to a significant reduction in the upstream collisional areas (see figures \ref{Relative_trajectories_for_cases_VII_and_VIII}$b$,$d$ and \ref{Relative_trajectories_for_cases_I_and_II}$d$).

\begin{figure}
\centering
\includegraphics[width=1.0\textwidth]{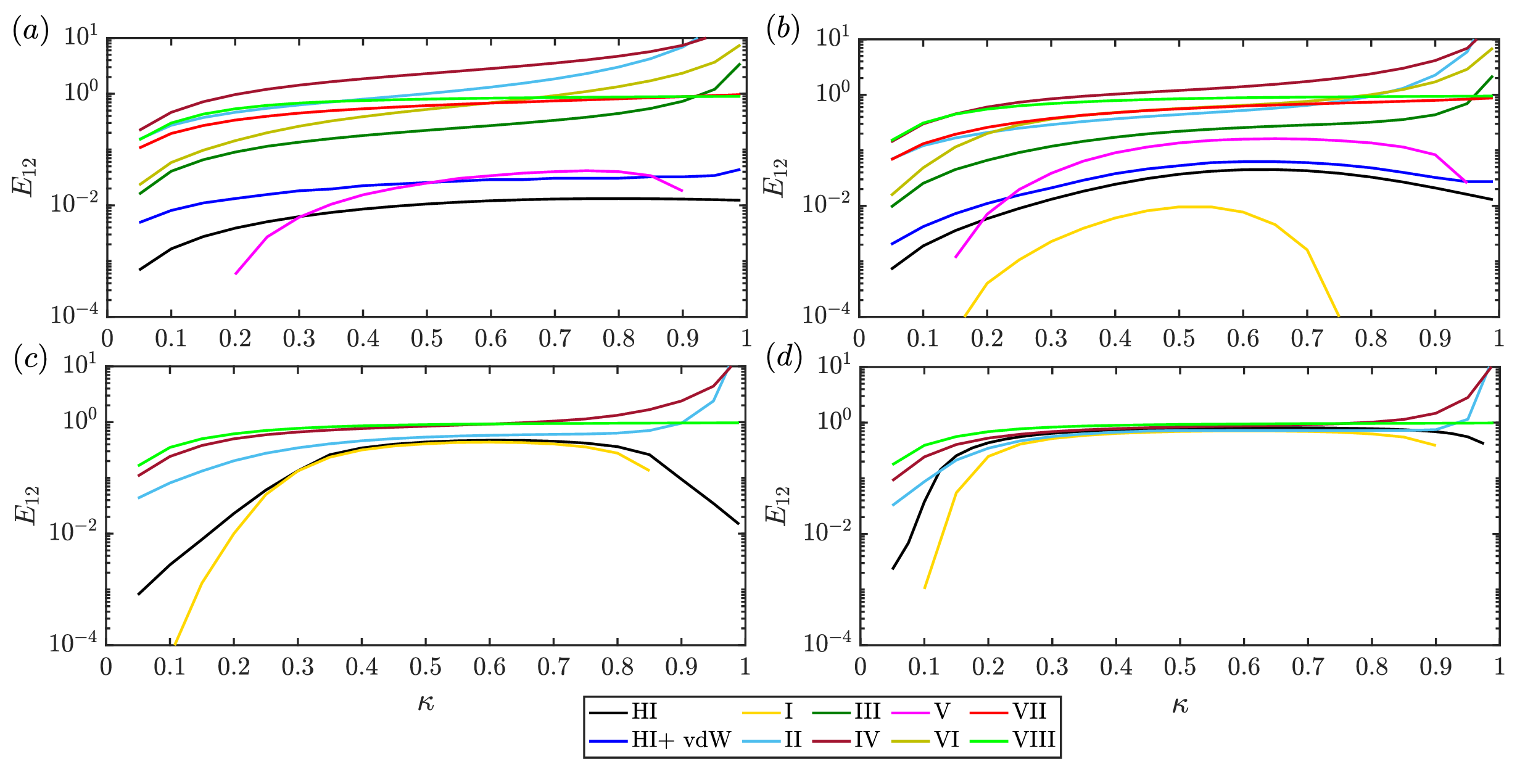}
\caption{Collision efficiency $E_{12}$ as a function of the droplet size ratio $\kappa=a_2/a_1$ for two droplets settling in still air under different charge and electric field conditions. Panels $(a)$--$(d)$ correspond to radii of the larger droplet of $a_1=10$, $20$, $30$ and $40\,\text{\textmu}\mathrm{m}$, respectively. Black curves show the collision efficiencies obtained with hydrodynamic interactions alone (HI) and agree with the results of \citet{ababaei2023collision}. Blue curves show the collision efficiency including the van der Waals attraction (HI+vdW). In the absence of an applied field, cases I and II correspond to like-charged and oppositely charged droplets, respectively. In a vertical field, cases III--IV, V--VI, and VII--VIII correspond to uncharged, like-charged, and oppositely charged droplet pairs, respectively. Within each pair of cases, the field strength is $5\times10^4\,\mathrm{V\,m^{-1}}$ in the first case and $3\times10^5\,\mathrm{V\,m^{-1}}$ in the second (Table~\ref{Different_cases_of_electric_charg_and_field_of_droplets}). Because several curves cluster at intermediate $\kappa$ and are difficult to distinguish, panels $(c)$ and $(d)$ show only the HI reference and cases I, II, IV, and VIII. As $\kappa$ approaches unity, $E_{12}$ increases sharply in cases II and IV and remains close to unity in case VIII.}
\label{Collision_efficiency_for_a1_10_20_30_40}
\end{figure}

Having shown the influence of electrostatic interactions on relative trajectories, we now examine the effects of these interactions on collision efficiency. Figures \ref{Collision_efficiency_for_a1_10_20_30_40}$a$-$d$ present the collision efficiency computed using the trajectory analysis discussed above, for various combinations of droplet surface charges and electric field strengths. The four panels correspond to cases with four different radii of the larger droplets: $a_1=10,20,30,40$ \textmu m. The horizontal axis is the radius of the smaller droplet $a_2=\kappa a_1$, plotted as the droplet size ratio $\kappa$. We begin by briefly discussing the collision efficiency in the absence of electrostatic interactions. The black solid lines in each panel give the variation of collision efficiency with size ratio when droplet pairs are subjected to only gravitational and hydrodynamic forces. We find that collision efficiency peaks at intermediate size ratios. For small $\kappa$ values, the smaller droplet experiences minimal inertia, causing it to follow the fluid streamlines and move around the larger droplet without colliding, unless it happens to follow a streamline very close to the larger droplet. As $\kappa$ increases, along with the absolute sizes of the droplets, the inertia of the droplets also rises, thereby facilitating more collisions. However, when the droplets are comparable in size, their relative velocity decreases, resulting in a slower evolution of the squeeze film between them and, consequently, larger interdroplet viscous stresses, leading to lower collision efficiency. In the presence of attractive van der Waals forces, droplet pairs remain in close proximity for a longer duration, allowing these forces to pull the droplets into contact rapidly and resulting in higher collision efficiencies, as shown by the blue solid lines in figures \ref{Collision_efficiency_for_a1_10_20_30_40}$a$-$b$. We see these two effects in competition as $a_1$ increases from $10$ to $20\,\text{\textmu}\mathrm{m}$; although the van der Waals attraction is a volume-volume interaction, as the size of the larger droplet increases, so too does the viscous resistance discussed above.  Therefore, the percentage increase in the collision efficiency relative to hydrodynamic interactions alone is larger when $a_1=10\,\text{\textmu}\mathrm{m}$. Our computed values for collision efficiency and their trends with $\kappa$ are consistent with the findings reported by \citet{ababaei2023collision} and \citet{rother2022gravitational}.


We will use collision efficiencies that exclude van der Waals and electrostatic interactions as a reference for comparison with those that include electrostatic interactions. Initial comparisons of collision efficiencies shown in different panels of figure \ref{Collision_efficiency_for_a1_10_20_30_40} suggest that the significance of electrostatic effects diminishes as the larger droplet radius increases, primarily due to the prevailing influence of droplet inertia. The collision efficiencies of two positively charged sedimenting droplets, in the absence of an electric field (represented as line I), are consistently lower than those of uncharged droplets due to the repulsive forces acting between them. Because a negatively charged droplet pair in the absence of an external electric field yields results identical to those represented by line I, we do not present them separately. For smaller droplets, such as when $a_1 = 10$ \textmu m, the collision efficiency of the positively charged pair is either zero or so low that it falls below the scale shown in figure \ref{Collision_efficiency_for_a1_10_20_30_40}$a$. This is primarily due to the dominance of the repulsive forces at large droplet separation distances. For case I, the strengths of near-field electrostatic attraction forces are greater for intermediate size ratios, and therefore, collision efficiency values are also higher for these intermediate size ratios (see line I in figures \ref{Collision_efficiency_for_a1_10_20_30_40}$b$-$c$). It is important to note that, as reported by \citet{khain2004rain} and \citet{patra2023collision}, when the charge ratios between like-charged droplets are high, the strong near-field electrostatic attraction resulting from intense mutual polarization can lead to higher collision efficiencies compared to those of uncharged pairs. However, the current study does not consider scenarios with high droplet charge ratios. In the absence of an external electric field, droplets that carry opposite charges invariably experience enhanced collision efficiency due to the attractive forces between them (see line II). This enhancement is most pronounced for smaller droplets. As the electrostatic attraction force between oppositely charged droplets increases as the interdroplet size difference decreases, the collision efficiency increases with higher $\kappa$ values. For larger droplets, especially those with $a_1 > 40$ \textmu m, the effect of droplet charges on collision efficiency becomes negligible due to the dominant influence of droplet inertia. 

As shown by lines III and IV in figures \ref{Collision_efficiency_for_a1_10_20_30_40}$a$-$d$, a vertical external electric field invariably enhances the collision efficiency of two uncharged sedimenting cloud droplets. This enhancement occurs because the electric field induces opposite charges on the adjacent surfaces of the droplets, thereby generating attractive electrostatic forces between them. We find that collision efficiency increases with the size ratio, a trend that reflects the strengthening of electric-field-induced forces as the size disparity between the two droplets reduces. As expected, strong electric fields significantly enhance collision efficiencies for smaller droplets. Notably, in cases where the larger droplet sizes are $10$ and $20$ \textmu m, the collision efficiency increases by an order of magnitude when the electric field strength rises from $5 \times 10^4$ Vm$^{-1}$ to $3 \times 10^5$ Vm$^{-1}$.

As  discussed previously, an applied electric field counteracts the repulsive interaction that occurs between like-charged droplets. Consequently, the collision efficiencies of like-charged droplet pairs subjected to a vertical electric field, shown by lines V and VI in figure \ref{Collision_efficiency_for_a1_10_20_30_40}, are consistently higher than those obtained in the absence of an electric field. However, these efficiencies remain lower than those of electrically neutral droplet pairs of the identical sizes subjected to an electric field of equal strength. In case V, where the applied field is relatively weak ($E_0=5 \times 10^4$ Vm$^{-1}$), the repulsive electrostatic forces, resulting directly from the surface charges of the two like-charged droplets, continue to dominate the electric field effects. Therefore, the relationship between collision efficiency and size ratio in case V is qualitative similar to that of case I. By contrast, in case VI where the field strength is higher ($E_0=3 \times 10^5$ Vm$^{-1}$), the electric-field-induced attractive forces surpass the repulsive forces, resulting in a monotonic increase in collision efficiency with size ratio. The collision efficiency of two oppositely charged sedimenting droplets subjected to a vertical electric field, shown by lines VII and VIII in figure \ref{Collision_efficiency_for_a1_10_20_30_40}, increases with increasing droplet size ratio. A comparison of cases VII and VIII with cases II, III, and IV shows that the combined action of the applied electric field and the surface charges of two oppositely charged droplets can either enhance or suppress collisions relative to scenarios in which surface-charge effects and electric-field effects act independently; the outcome depends on the droplet sizes and the field strength. Nonetheless, the collision efficiencies in cases VII and VIII are always higher than those obtained in the absence of any electrostatic interactions.

\section{Droplet size distribution evolution}\label{sec:DSD_evolution}

The evolution of the droplet size distribution (DSD) due to collisions is described by the stochastic collection equation (SCE) \citep{pruppacher1997microphysics}. Here, electric charges and external electric fields are taken into consideration for droplet-droplet collisions. The droplet distribution function 
depends on droplet mass $m$ (or radius $a$) and electric charge $q$, so that the SCE is expressed as
\begin{eqnarray}
    \frac{\partial n(m,q,t)}{\partial t}=\int_{m_0}^{m/2}\Big[\int_{-\infty}^{\infty}K(m',q';m-m',q-q')n(m',q',t) \nonumber \\ \cdot n(m-m',q-q',t)dq'\Big]dm' \nonumber \\ -\int_{m_0}^{\infty}\Big[\int_{-\infty}^{\infty}K(m,m';q,q')n(m,q,t)n(m',q',t)dq'\Big]dm'.
\label{DSD_equation}
\end{eqnarray}
In \eqref{DSD_equation}, $n(m,q,t)$ is the droplet number distribution function
at time $t$ and $K(m',q';m-m',q-q')$ is the collision kernel, which represents the rate at which a droplets of mass $m'$ and charge $q'$ collide with droplets of mass $m-m'$ and charge $q-q'$ to form new droplets of mass $m$ and charge $q$. Here, $m_0$ is the mass of the smallest droplet involved in the collision process. The first integral (or ``gain integral'') on the right-hand side of Eq. \eqref{DSD_equation} describes the gain of droplets of mass $m$, while the second integral (or ``loss integral'') denotes the loss of droplets with mass $m$ due to collisions with other droplets of any size.

Many different numerical methods have been developed to solve the SCE \citep[See e.g.,][]{berry1974analysis,bott1998flux,shima2009super}. Here, we solve the SCE using a flux-based method developed by \citet{bott1998flux}, although since our collision kernel depends on both the droplet sizes and charges, the procedure of solving the SCE is more complicated than that given therein, where the collision kernel depended solely on droplet sizes. Therefore, we introduce logarithmically equidistant two-dimensional mass charge grids. We consider $1000$ mass bins that describe the droplet size distribution with droplet radii ranging from $10^{-6}$ m to $2500 \times 10^{-6}$ m. In each radius bin, droplets may have different amounts and different signs of charges. For the bin of radius $a$, droplet charge ranges from $-2 \times 3.33 \times 10^{-6} a^2$ C to $2 \times 3.33 \times 10^{-6} a^2$ C. This means that smaller droplets have a smaller range of charge. We divide the charge range into 35 bins, with the center bin having zero charge, and the negative charge bins are symmetric to the positive charge bins. During the evolution of the DSD, the radius and charge of colliding droplets have large variability. Therefore, for the size bins and charge bins described above, we compute a large matrix of kernels $K(a_1,q_1,a_2,q_2)$ in advance as a lookup table for solving the SCE. We have seen that the effect of electrostatic interactions on collision efficiency decreases significantly as the radius of the larger droplet in the pair exceeds $40$ \textmu m. Thus, for $a_1 > 40$ \textmu m, we neglect the electrostatic effects on collision efficiency. Notably, for larger droplets, both droplet Reynolds number ($Re_d$) and Stokes number ($St$) will play a significant role in the collision dynamics. In our simulations, for $a_1 > 40$ \textmu m, we use the collision efficiency data given by \citet{pinsky2001collision}.

\begin{figure}
\centering
\includegraphics[width=1.0\textwidth]{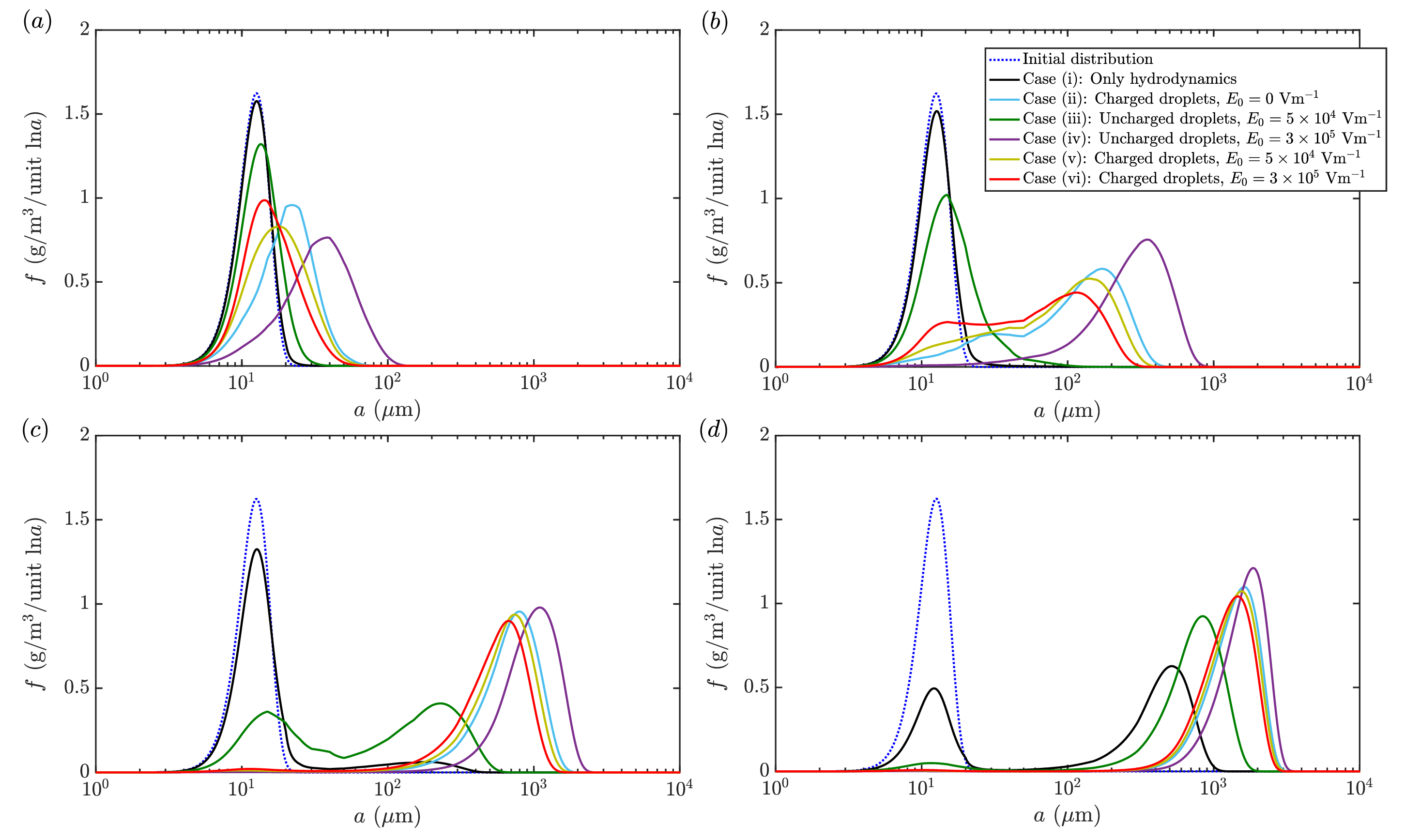}
\caption{The droplet mass distribution function after $(a)$ $10$ min, $(b)$ $20$ min, $(c)$ $30$ min and $(d)$ $40$ min for different cases shown in the inset (and described below) when $L=1$ gm$^{-3}$ and $\overline{a}=10$ \textmu m. The color legend for $(a)$, $(c)$ and $(d)$ is the same as for $(b)$. Blue dotted lines are the initial droplet mass distribution function for all cases. In the main text we refer to the six cases in the inset as (i)--(vi) as follows:  (i) droplets with only hydrodynamic interactions, (ii) charged droplets without an electric field, (iii) uncharged droplets with a vertical electric field of strength $5 \times 10^4$ Vm$^{-1}$, (iv) uncharged droplets with a vertical electric field of strength $3 \times 10^5$ Vm$^{-1}$, (v) charged droplets with a vertical electric field of strength $5 \times 10^4$ Vm$^{-1}$, and (vi) charged droplets with a vertical electric field of strength $5 \times 10^4$ Vm$^{-1}$.}
\label{Size_distribution_after_10_20_30_and_40_min}
\end{figure}

Because the masses and charges of droplets formed after collisions do not fall into any existing bins, it is necessary to remap these newly formed droplets into the regular mass and charge bins. The remapping process involves two sequential steps: mass remapping followed by charge remapping. Following the procedure described in \citet{khain2004rain}, we first distribute the droplets among the radius bins in proportions that simultaneously conserve total mass and droplet number. Within each radius bin, we then distribute the droplets to the corresponding charge bins in a manner that conserves both total charge and droplet number.


We assume that the initial droplet number distribution is exponential in droplet mass,
\begin{equation}
    n(m,0) = \frac{L}{\overline{m}^{\,2}}
    \exp\!\left(-\frac{m}{\overline{m}}\right),
    \label{Initial_DSD}
\end{equation}
where $n(m,0)\,\mathrm{d}m$ is the initial number density of droplets with masses between $m$ and $m+\mathrm{d}m$, $L$ is the cloud liquid water content, and $\overline{m}$ is the mean droplet mass. The corresponding mean volume radius is $\overline{a} = [3\overline{m}/(4\pi\rho_d)]^{1/3}$, and the initial total droplet number density is $L/\overline{m}$. For the simulations with charged droplets, we follow \citet{guo2021enhancement} and prescribe a Gaussian distribution of charge within each mass bin, with zero mean and a bin-dependent standard deviation. Each mass bin is therefore initially neutral on average, although individual droplets can carry either positive or negative charge.


In our simulation, we set the liquid water content to $1$ gm$^{-3}$, a value representative of typical warm cumulus clouds. We chose an initial size distribution with a mean droplet radius of $10$ \textmu m, which yields an initial number concentration of $238.73$ cm$^{-3}$. The integration time step chosen in these simulations was $1$ s. Figures \ref{Size_distribution_after_10_20_30_and_40_min}$a$-$d$ show the simulated evolution of the DSD at $10$, $20$, $30$, and $40$ min, respectively. In each figure panel, the horizontal axis is the droplet radius in \textmu m and the vertical axis is the droplet mass distribution function $f(\ln a)$. This function $f(\ln a)$ is related to $n(m)$ as $f(\ln a)\,d\ln a = m\,n(m)\,dm$. The blue dotted line in each panel is the initial size distribution. At $10$-minute intervals we compare the DSDs for six different cases; (i) droplets with only hydrodynamic interactions, (ii) charged droplets without an electric field, (iii) uncharged droplets with a vertical electric field of strength $5 \times 10^4$ Vm$^{-1}$, (iv) uncharged droplets with a vertical electric field of strength $3 \times 10^5$ Vm$^{-1}$, (v) charged droplets with a vertical electric field of strength $5 \times 10^4$ Vm$^{-1}$, and (vi) charged droplets with a vertical electric field of strength $5 \times 10^4$ Vm$^{-1}$. At each 10-minute interval we find that the DSDs with different electric conditions differ significantly from one another. At $t=10$ min, all DSDs show uni-modal distribution. For cases (ii), (v), and (vi), second modes in their size distribution gradually form as droplets undergo the collision–coalescence process from $10$ min to $20$ min. Due to high collision rates, the second peak for case (iv) (not shown here) formed and disappeared between 10 and 20 minutes. At $t=30$ min, the second mode is quite obvious for case (iii), whereas it just starts to grow for case (i). Except for cases (i) and (iii), all other distributions are uni-modal, showing relatively faster evolution due to strong electrostatic interactions. At $t=40$ min, the modal radius of the cases with electrostatic effects is much higher than that of the second peak of the case without any electrostatic effects. Thus, from these distributions, we conclude that electrostatic interactions in strongly electrified clouds can trigger rain formation relatively quickly by enhancing the droplet collision process.


\section{Conclusions}\label{sec:Conclusions}

We have quantitatively assessed the influence of electric charges and atmospheric electric fields on cloud droplets settling under gravity in a quiescent atmosphere. Our analysis employs exact hydrodynamic and electrostatic interactions in the calculation of collision rates. We build upon recent analytical formulations of electrostatic forces at close separation distances, demonstrating that these forces can overcome lubrication resistance and facilitate surface-to-surface contact within a finite time. Most importantly, we have included the effects of internal circulations and interfacial mobilities, which are crucial for accurately modeling cloud droplet collisions. By examining the relative trajectories of the droplet pairs, we have provided a comprehensive framework for calculating the upstream interception area, a vital parameter in evaluating the collision rate.

We have demonstrated that the interplay between droplet inertia, electric charge, and the applied electric field significantly influences droplet–droplet collision outcomes. Our results quantitatively show how collision efficiency varies with the droplet size ratio across different combinations of droplet charge and electric field strength. We find that collision efficiency increases for oppositely charged pairs or for uncharged/charged droplets subjected to a strong vertical electric field. This increase in collision efficiency is particularly significant for small cloud droplets. Using the collision efficiencies we obtained, we solve the stochastic collection equation to simulate the evolution of the DSD in clouds under the influence of electrostatic effects. Our findings suggest that for a cloud with initial $\overline{a}=10$ \textmu m, electrostatic interactions could significantly enhance droplet collision, thereby substantially intensifying the collision–coalescence process and facilitating the onset of raindrop formation.

A crucial factor not addressed in this study is atmospheric turbulence, which plays a key role in the growth of cloud droplets and the initiation of rainfall \citep{shaw2003particle}. In the size-gap regime, turbulence enhances collision rates through (a) increased radial relative velocities between droplet pairs \citep{saffman1956collision,falkovich2007sling} and (b) preferential concentration of inertial droplets in straining regions of the flow \citep{sundaram1997collision,chun2005clustering}. Additionally, the velocity perturbations induced by droplet motion further influence collision rates under realistic conditions \citep{pinsky2007collisions}. Recent direct numerical simulations coupled with Lagrangian particle tracking have demonstrated that turbulence and droplet aerodynamic interactions significantly broaden the droplet size distribution \citep{chen2018turbulence,michel2023influence}. A recent study by \citet{patra2025electric} on droplets in a background linear flow with an electric field reported results consistent with the present study's findings for pure gravitational settling — showing that electric fields can significantly enhance droplet collisions. However, there are currently no theoretical predictions for collision efficiencies of inertial droplets under the combined effects of turbulence, gravity, and an external electric field. Future studies should aim to incorporate these factors into collision rate calculations, including a rigorous treatment of short-range electric-field-induced forces.  

Here we have focused exclusively on warm clouds, where droplet collisions govern the formation of raindrops. However, mixed-phase clouds contain both supercooled droplets and ice crystals, and their interactions play a fundamental role in the formation of precipitation. For instance, snow aggregates form through collisions between ice crystals, while graupel growth occurs via riming - when settling ice crystals collide with supercooled droplets in turbulent conditions \citep{pruppacher1997microphysics}. While spherical droplet collisions are well understood, ice crystal collisions remain less explored due to their anisotropic shapes and variable settling orientations. Recent studies \citep{jucha2018settling,sheikh2022colliding} have examined turbulence and gravitational effects on ice crystal collisions, while others have explored collisions between ice crystals and supercooled droplets \citep{naso2018collision,jost2019effect,sheikh2024effect}. However, these studies rely on the ghost collision approximation, which neglects hydrodynamic and electrostatic interactions between colliding hydrometeors. Given that electrostatic forces can become significant at close distances, they can substantially influence the outcomes of ice crystal collisions. Recent work by \citet{joshi2025electrostatic} has quantified the electrostatic forces and torques between charged anisotropic particles, demonstrating that electrostatic torques can induce the preferential alignment of ice crystals, thereby affecting their interactions. Future research should incorporate these electrostatic effects into collision rate calculations, considering both translational and rotational dynamics to achieve a more comprehensive understanding of hydrometeor interactions due to electric effects.

\vspace{5mm}
\textbf{Acknowledgments.} P.P. and J.S.W.  gratefully acknowledge support from the Swedish Research Council under Grant No.638-2013-9243.

\vspace{5mm}
\textbf{Declaration of interests.} The authors report no conflict of interest.

\vspace{5mm}
\textbf{Data availability.} The data that support the findings of this article are not publicly available. The data are available from the corresponding author upon reasonable request.

\appendix

\section{The force coefficients associated with the electrostatic interactions}\label{appA}

The force coefficients in Eq. (\ref{Electrostatic_force_on_droplet_2}) are complicated series expressions that require numerical evaluation. As the droplet separation distance decreases, the number of terms required for the series to achieve convergence increases significantly. Consequently, the computational cost of evaluating these series at small separation distances becomes prohibitively expensive. Therefore, employing the equivalent analytical expressions for these series in the near field becomes imperative. In order to make this paper reasonably self contained, we collect here the analytical formulations of these coefficients as documented in the existing literature.

Only coefficients $F_5$, $F_6$, and $F_7$ are necessary for characterizing forces that charged droplets experience in the absence of an externally applied electric field. The pioneering work by \citet{russell1909coefficients}, followed by subsequent contributions from \citet{lekner2012electrostatics}, led to the derivation of an analytical expression for the electrostatic force in the lubrication region between two charged conducting spheres. These studies described the electrostatic interactions between charged conducting spheres through the capacitance coefficients of the system. For a pair of spherical conductors, the linear relationship between their respective charges $q_i$ and electrical potential $\mathcal{V}_i$ ($i=1,2$) is
\begin{equation}
     \begin{bmatrix}
     q_{1} \\
     q_{2}
     \end{bmatrix}
     =
    \begin{bmatrix}
    C_{11} & C_{12} \\
    C_{21} & C_{22} 
    \end{bmatrix}
    \begin{bmatrix}
    \mathcal{V}_{1} \\
    \mathcal{V}_{2}
    \end{bmatrix}
    ,
\end{equation}
where $C_{11}$, $C_{22}$, $C_{12}$, and $C_{21}$ denote the capacitance coefficients of the system. The electrostatic potential energy $W$ of this system is given by:
\begin{equation}
    W = \frac{1}{2}\left(q_1 \mathcal{V}_1 + q_2 \mathcal{V}_2\right) = \frac{q_2^2 C_{11} - 2 q_1 q_2 C_{12} + q_1^2 C_{22}}{2\left(C_{22} C_{11} - C_{21}^2\right)}. \label{Electric_potential_energy}
\end{equation}

We can calculate $W$ for specified charges and potentials if we know the capacitance coefficients. When the dimensionless surface-to-surface distance is small, $\xi \ll 1$, these capacitance coefficients can be expressed as:
\begin{subeqnarray}
   C_{11} & = & 4\pi\epsilon_0 \frac{a_1 a_2}{a_1 + a_2}\left\{\frac{1}{2}\ln\left[\frac{4\lambda(1-\lambda)}{\xi}\right]-\psi(\lambda)+O(\xi)\right\}, \label{C11_expression_small_xi} \\ [3pt]
   C_{22} & = & 4\pi\epsilon_0 \frac{a_1 a_2}{a_1 + a_2} \left\{\frac{1}{2}\ln\left[\frac{4\lambda(1-\lambda)}{\xi}\right]-\psi(1-\lambda)+O(\xi)\right\} \quad \textrm{and} \label{C22_expression_small_xi} \\ [3pt]
   C_{12} & = & C_{21} = -4\pi\epsilon_0 \frac{a_1 a_2}{a_1 + a_2} \left\{\frac{1}{2}\ln\left[\frac{4\lambda(1-\lambda)}{\xi}\right]+\gamma+O(\xi)\right\}, \label{C12_expression_small_xi}
\end{subeqnarray}
where $\lambda=\kappa/(1+\kappa)$, $\psi$ represents the digamma function, and $\gamma=-\psi(1)\approx0.57722$ is the Euler–Mascheroni constant. After substituting the expressions for $C_{11}$, $C_{22}$, and $C_{12}$ into (\ref{Electric_potential_energy}), one can derive the expression for the electrostatic interaction force by evaluating the term $-\partial W/\partial r=-[2/(a_1+a_2)]\partial W/\partial \xi$. A comparison of the expression of the force acting on droplet $2$, derived through this methodology, with the expression provided in Eq. (\ref{Electrostatic_force_on_droplet_2}) under the condition that $E_0=0$, yields
\begin{subeqnarray}
   F_5 & \sim & \frac{2\kappa}{\xi} \left[\frac{\gamma+\psi(1-\lambda)}{\left[2\gamma + \psi(\lambda) + \psi(1-\lambda) \right]\ln \left[4\lambda(1-\lambda)/\xi\right] + 2\gamma^2 - 2 \psi(\lambda) \psi(1-\lambda)}\right]^2 + O(1), \label{F5_expression_small_xi} \\ [3pt]
   F_6 & \sim & -\frac{4\kappa}{\xi} \left[\frac{(\gamma+\psi(\lambda))(\gamma+\psi(1-\lambda))}{\left\{\left[2\gamma + \psi(\lambda) + \psi(1-\lambda) \right]\ln \left[4\lambda(1-\lambda)/\xi\right] + 2\gamma^2 - 2 \psi(\lambda) \psi(1-\lambda)\right\}^2}\right] + O(1) \quad \textrm{and} \label{F6_expression_small_xi} \\ [3pt]
   F_7 & \sim & \frac{2\kappa}{\xi} \left[\frac{\gamma+\psi(\lambda)}{\left[2\gamma + \psi(\lambda) + \psi(1-\lambda) \right]\ln \left[4\lambda(1-\lambda)/\xi\right] + 2\gamma^2 - 2 \psi(\lambda) \psi(1-\lambda)}\right]^2 + O(1). \label{F7_expression_small_xi}
\end{subeqnarray}
\citet{lekner2012electrostatics} also showed that two charged conducting spheres always exhibit an attractive interaction at short distances, even when the polarity of electrical charges on both spheres is the same, except for specific charge ratio values that the spheres would attain by bringing them into contact. Notably, the near-field scaling behaviors of $F_5$, $F_6$, and $F_7$ imply that the attractive force has an $O(\xi^{-1}[\ln\xi]^{-2})$ singularity, which diverges as the separation distance approaches zero.

In scenarios where $r \gg (a_1+a_2)$, the capacitance coefficients can be expressed as inverse powers of $r$, as given by \citet{lekner2012electrostatics}:  
\begin{subeqnarray}
   C_{11} & = & 4\pi\epsilon_0\left[ a_1 + \frac{a_1^2 a_2}{r^2} + \frac{a_1^2 a_2^2 (a_1+a_2)}{r^4} + \frac{a_1^2 a_2^2 (a_1^3 + a_1^2 a_2 + 2 a_1 a_2^2 + a_2^3)}{r^6} + O(r^{-8}) \right], \label{C11_expression_large_xi} \\ [3pt]
   C_{22} & = & 4\pi\epsilon_0 \left[ a_2 + \frac{a_1 a_2^2}{r^2} + \frac{a_1^2 a_2^2 (a_1+a_2)}{r^4} + \frac{a_1^2 a_2^2 (a_1^3 + 2 a_1^2 a_2 +  a_1 a_2^2 + a_2^3)}{r^6} + O(r^{-8})\right] \quad \textrm{and} \label{C22_expression_large_xi} \\ [3pt]
   C_{12} & = & -4\pi\epsilon_0\left[\frac{a_1 a_2}{r} + \frac{a_1^2 a_2^2}{r^3} + \frac{a_1^2 a_2^2 (a_1^2 + a_1 a_2 +a_2^2)}{r^5} + \frac{a_1^2 a_2^2 (a_1^2 + a_1 a_2 +a_2^2)^2}{r^7} + O(r^{-9})\right]. \label{C12_expression_large_xi}
\end{subeqnarray}
Consequently, the far-field analytical expressions for $F_5$, $F_6$, and $F_7$, derived by using (\ref{C11_expression_large_xi}), can be succinctly stated as follows:
\begin{subeqnarray}
   F_5 & = & \frac{64\lambda^5}{(2+\xi)^5} + \frac{384\lambda^7}{(2+\xi)^7} + \frac{2048\lambda^9}{(2+\xi)^9} + O(\xi^{-11}), \label{F5_expression_large_xi} \\ [3pt]
   F_6 & = & \frac{4\lambda^2}{(2+\xi)^2} + \frac{3584\kappa^5(1-\lambda)^8}{(2+\xi)^8} + \frac{27648\kappa^5(1+\kappa^2)(1-\lambda)^{10}}{(2+\xi)^{10}} + O(\xi^{-12}) \quad \textrm{and} \label{F6_expression_large_xi} \\ [3pt]
   F_7 & = & \frac{64 \kappa^2 (1-\lambda)^5}{(2+\xi)^5} + \frac{384 \kappa^2(1-\lambda)^7}{(2+\xi)^7} + \frac{2048\kappa^2(1-\lambda)^9}{(2+\xi)^9} + O(\xi^{-11}). \label{F7_expression_large_xi}
\end{subeqnarray}

To calculate the electrostatic forces acting on two uncharged conducting spheres in an external electric field, the only coefficients required are $F_1$, $F_2$, and $F_8$. The forces in this scenario can be expressed in terms of the longitudinal ($\alpha_l$) and transverse ($\alpha_t$) polarizabilities of the system comprising two spherical conductors \citep{lekner2013forces}. The mathematical expressions for these polarizabilities involve infinite series that exhibit slow convergence, particularly at small sphere-sphere separation distances. In order to circumvent this convergence issue, \citet{lekner2011polarizability} used the Abel-Plana formula to transform these series sums into integrals. To evaluate electric field-induced forces in the lubrication regime, we employ the analytical solutions for these polarizabilities in the close-approach limit given by \citet{lekner2011polarizability}. A comprehensive discussion on force and torque calculations derived using the polarizability tensor is available in \citet{lekner2013forces}. By comparing the expression of the force acting on droplet $2$, derived in terms of polarizabilities, with the expression from equation (\ref{Electrostatic_force_on_droplet_2}) for the case when $q_1=q_2=0$, we obtain
\refstepcounter{equation}
$$
  F_1 = -\frac{1}{2 a_2^2} \frac{\partial \alpha_l}{\partial r}, \quad
  F_2 = -\frac{1}{2 a_2^2} \frac{\partial \alpha_t}{\partial r} \quad \textrm{and} \quad
  F_8 = -\frac{1}{2 a_2^2} \frac{\left(\alpha_l-\alpha_t\right)}{r}.
  \eqno{(\theequation{\mathit{a},\mathit{b},\mathit{c}})} \label{Expresions_for_F1_F2_and_F8}
$$
The analytical expressions of $\alpha_l$ and $\alpha_t$ for small separation distances are given in \citet{lekner2011polarizability} as follows:
\begin{subeqnarray}
\alpha_l & = &  a_2^3(1-\lambda)^3\Bigg[ 4\zeta(3)- \psi''(\lambda) -\psi''(1-\lambda) + \frac{\mathcal{M}(\xi,\lambda)}{\mathcal{N}(\xi,\lambda)}\Bigg] \quad \textrm{and}\label{alpha_l_small_xi} \\ [3pt]
\alpha_t & = &  a_2^3(1-\lambda)^3\Big[\mathcal{Z}(\lambda)+\frac{\xi}{2\lambda(1-\lambda)}\mathcal{T}(\lambda)\Big], \label{alpha_t_small_xi}
\end{subeqnarray}
where $\zeta(3)\approx1.2020569$ and the prime denotes an ordinary derivative. The expressions for the other variables involved in (\ref{alpha_l_small_xi}) are
\begin{subeqnarray}
    \mathcal{M}(\xi,\lambda) & = & \frac{1}{2}[\psi'(\lambda)-\psi'(1-\lambda)]^2 \ln\left({\frac{4\lambda(1-\lambda)}{\xi}}\right) -\frac{\pi^4}{36}\mathcal{F}(\lambda)-\frac{\pi^2}{3}\mathcal{G}(\lambda)-\mathcal{H}(\lambda), \\ [3pt] 
    \mathcal{N}(\xi,\lambda) & = & \frac{1}{2} \mathcal{F}(\lambda) \ln\left(4\lambda(1-\lambda)\right)+\gamma^2-\psi(\lambda)\psi(1-\lambda) - \frac{\mathcal{F}(\lambda)}{2} \ln{\xi}, \\ [3pt]
   \mathcal{Z}(\lambda) & = & -2\zeta(3)-\frac{1}{2}\psi''(\lambda)-\frac{1}{2}\psi''(1-\lambda) \quad \textrm{and} \\ [3pt]
  \mathcal{T}(\lambda) & = & \mathcal{F}(\lambda) + (1-3\lambda+3\lambda^2) \mathcal{Z}(\lambda) + \frac{1}{6} \lambda(1-\lambda)(2\lambda-1)[\psi'''(\lambda)-\psi'''(1-\lambda)], 
\end{subeqnarray}
where
\begin{subeqnarray}
   \mathcal{F}(\lambda) & = & \psi(\lambda)+\psi(1-\lambda)+2\gamma , \\ [3pt]
   \mathcal{G}(\lambda) & = & \psi(\lambda)\psi'(1-\lambda)+\psi'(\lambda)\psi(1-\lambda) +\gamma[\psi'(\lambda)+\psi'(1-\lambda)] \quad \textrm{and} \\ [3pt]
   \mathcal{H}(\lambda) & = & \psi(\lambda)\psi'(1-\lambda)^2+2\gamma\psi'(\lambda)\psi'(1-\lambda) +\psi'(\lambda)^2\psi(1-\lambda).
\end{subeqnarray}
Equations (\ref{alpha_l_small_xi}) are valid at close approach only, at best up to a separation equal to the smaller of the two radii. Mathematically, these equations are accurate up to $\xi \approx 0.2\lambda(1-\lambda)$ (see \citet{lekner2011polarizability}). Finally, the scaling of the force coefficients $F_1$, $F_2$, and $F_8$ for $\xi \ll 1$ are governed by 
\begin{subeqnarray}
   F_1 & \sim & \frac{(1-\lambda)\mathcal{C}(\lambda)}{2\xi \Bigg[\dfrac{1}{2} \mathcal{F}(\lambda) \ln\left(4\lambda(1-\lambda)\right)+\gamma^2-\psi(\lambda)\psi(1-\lambda) - \dfrac{\mathcal{F}(\lambda)}{2} \ln{\xi}\Bigg]^2}, \label{F1_expression_small_xi} \\ [3pt]
   F_2 & \sim & -\frac{(1-\lambda)^2}{2} \mathcal{T}(\lambda) \quad \textrm{and} \label{F2_expression_small_xi} \\ [3pt]
   F_8 & \sim & -\frac{\lambda(1-\lambda)^3}{\xi + 2}\Bigg[  \frac{\mathcal{M}(\xi,\lambda)}{\mathcal{N}(\xi,\lambda)}-\frac{\xi}{2\lambda(1-\lambda)}\mathcal{T}(\lambda) +8\zeta(3)+\mathcal{Z}(\lambda)\Bigg]. \label{F8_expression_small_xi}
\end{subeqnarray}
where 
\begin{equation}
   \mathcal{C}(\lambda)=\lambda(1-\lambda)^2 \Bigg[\Big(\psi'(\lambda)-\psi'(1-\lambda)\Big)^2 \Big(\gamma^2 -\psi(\lambda)\psi(1-\lambda)\Big) 
+\mathcal{F}(\lambda) \Big(\frac{\pi^4}{36}\mathcal{F}(\lambda)+\frac{\pi^2}{3}\mathcal{G}(\lambda)+H(\lambda)\Big)\Bigg]. 
\end{equation}
These asymptotic expressions of $F_1$, $F_2$, and $F_8$ imply that $F_1=O\left(\xi^{-1}[\ln\xi]^{-2}\right)$, and both $F_2$, and $F_8$ approach constant values as $\xi \rightarrow 0$. The scaling of $F_1$ in the lubrication regime implies that the electric-field-induced forces can overcome the lubrication resistance and allow finite time contact between the droplet surfaces.

The longitudinal and transverse polarizabilities for large droplet separation distances are provided in \citet{lekner2013forces} and are
\begin{subeqnarray}
\alpha_l & = & a_1^3+a_2^3+\frac{4 a_1^3 a_2^3}{r^3}+\frac{4 a_1^3 a_2^3(a_1^3+a_2^3)}{r^6}+O(r^{-8}) \quad \textrm{and} \label{alpha_l_large_xi} \\ [3pt]
\alpha_t & = & a_1^3+a_2^3-\frac{2 a_1^3 a_2^3}{r^3}+\frac{a_1^3 a_2^3(a_1^3+a_2^3)}{r^6}+O(r^{-8}). \label{alpha_t_large_xi}
\end{subeqnarray}
Using equations (\ref{alpha_l_large_xi}), the far-field expressions for the relevant force coefficients are given by
\begin{subeqnarray}
F_1 & = & \frac{ 96 \lambda(1-\lambda)^3}{ r^4} + O(r^{-7}),
\label{F1_expression_large_xi} \\ [3pt]
F_2 & = & -\frac{48 \lambda(1-\lambda)^3}{ r^4} + O(r^{-7}) \quad \textrm{and}  
\label{F2_expression_large_xi} \\ [3pt] 
F_8 & = & -\frac{48\lambda(1-\lambda)^3}{ r^4} + O(r^{-7}).
\label{F8_expression_large_xi}
\end{subeqnarray}

The coefficients $F_3$, $F_4$, $F_9$, and $F_{10}$ characterize the interrelated effects of surface charges in conjunction with the applied electric field.  
Despite our attempts, we have not found in the literature information on the near and far field scalings of these coefficients.

\bibliography{References} 
\end{document}